\documentclass[11pt]{report}
\usepackage{setspace}
\usepackage{graphicx}
\usepackage{epstopdf}
\usepackage{amsmath}%
\usepackage{amsfonts}%
\usepackage{amssymb}%
\usepackage{epsf}%
\usepackage{hyperref}%
\usepackage{eso-pic}
\usepackage{graphicx}
\AddToShipoutPicture*{%
\put(0,0){\includegraphics[width=\paperwidth]{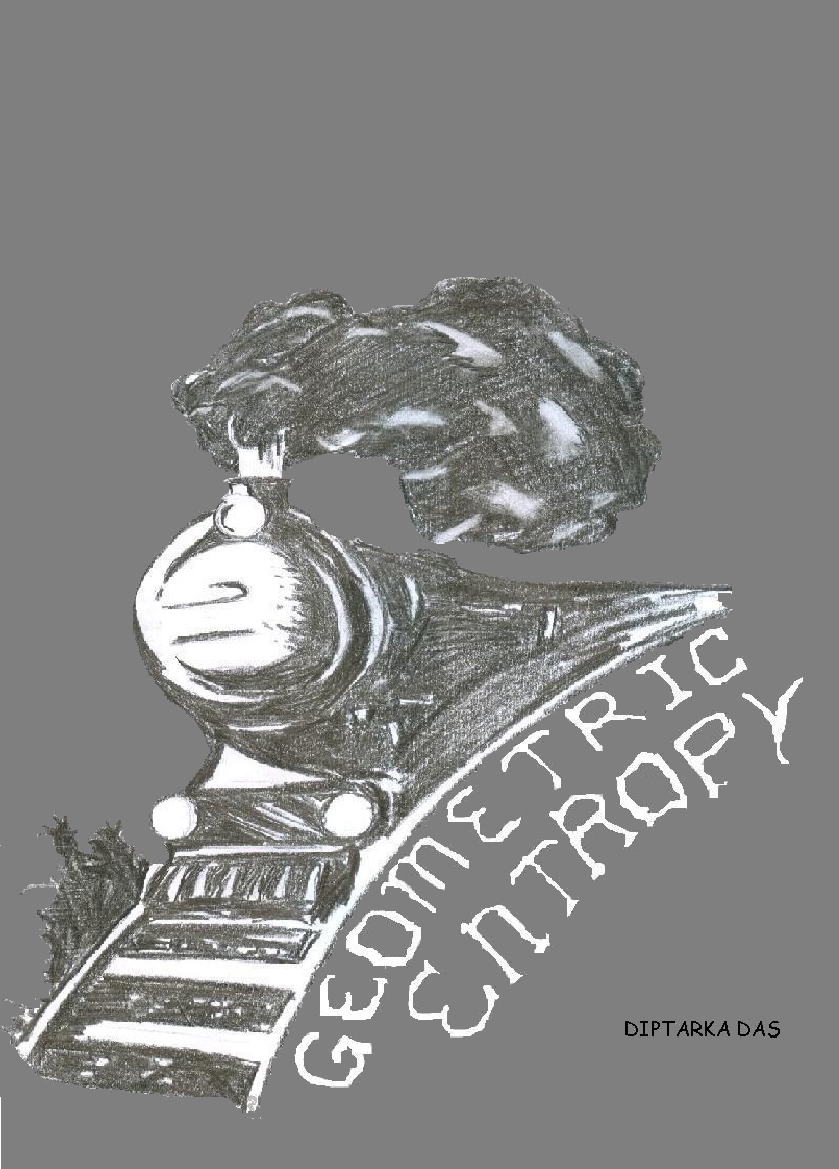}}%
}

\begin{document}
\newpage
\thispagestyle{empty}
\begin{center}
\includegraphics[width=0mm]{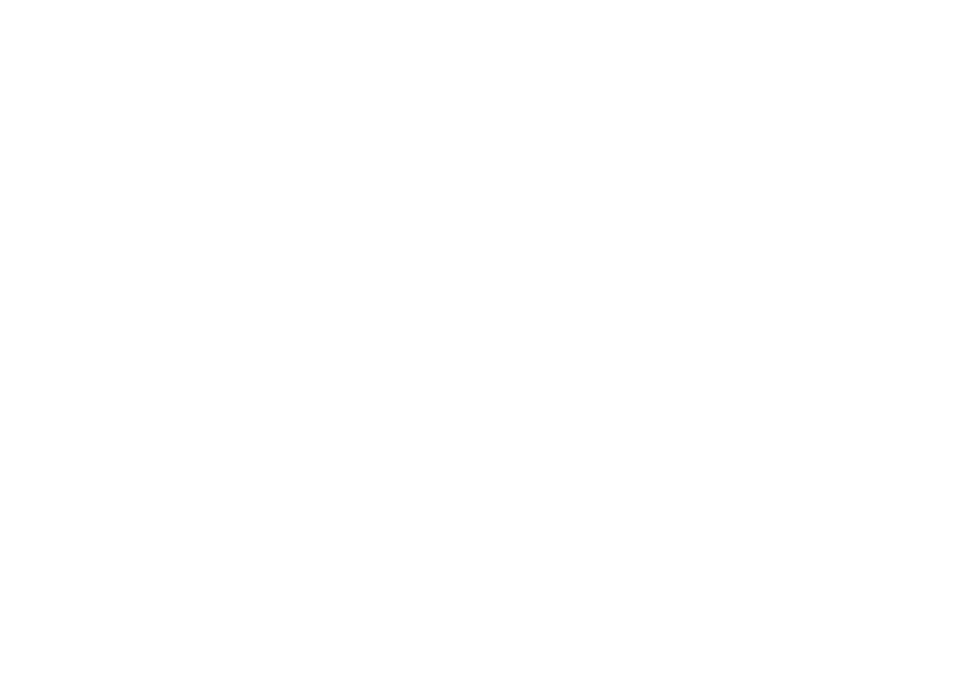}
\end{center}

\newcommand{\beq}{\begin{equation}}
\newcommand{\eeq}{\end{equation}}
\newcommand{\beg}{\begin{gathered}}
\newcommand{\eeg}{\end{gathered}}
\newcommand{\bal}{\begin{align}}
\newcommand{\eal}{\end{align}}
\newcommand{\bea}{\begin{eqnarray}}
\newcommand{\eea}{\end{eqnarray}}

\newcommand{\dee}{\partial}
\newcommand{\eps}{\mbox{{\boldmath $\epsilon$}}}
\newcommand{\Om}{{\bf \Omega}}
\newcommand{\om}{\mbox{{\boldmath $\omega$}}}
\newcommand{\de}{\delta}
\newcommand{\kap}{\kappa}
\newcommand{\cd}{\cdot}
\newcommand{\bTh}{{\bf \Theta}}
\newcommand{\al}{\alpha}
\newcommand{\be}{\beta}
\newcommand{\ga}{\gamma}
\newcommand{\bK}{{\bf K}}
\newcommand{\bS}{{\bf S}}
\newcommand{\bT}{{\bf T}}
\newcommand{\bL}{{\bf L}}
\newcommand{\bJ}{{\bf J}}
\newcommand{\bQ}{{\bf Q}}
\newcommand{\bE}{{\bf E}}
\newcommand{\mo}{{\mathcal {O}}}
\newcommand{\ma}{{\mathcal{A}}}
\newcommand{\bff}{{\bf f}}
\newcommand{\Liek}{{\cal L}_\xi}
\newcommand{\ph}{\varphi}
\def\d{{\mathrm{d}}}
\def\implies{\Rightarrow}
\def\Painleve{Painlev\'e}
\newpage
%begin a more systematic derivative naming
\newcommand{\del}{{\bf \nabla}}

\thispagestyle{empty}

\baselineskip=24pt

~\\

\begin{center}
\thispagestyle{empty}
{\bf {\Huge GEOMETRIC ENTROPY}}\\

\bigskip
\bigskip
\bigskip
\bigskip
\bigskip
\bigskip

Thesis\\
Submitted in partial fulfillment of the requirements of\\
BITS C421T/422T Thesis\\

\bigskip
\bigskip

By\\

\bigskip
\bigskip

{\bf {\Large Diptarka Das}}\footnote{diptarka.das@gmail.com}\\
ID No. 2005B5A3578P\\

\bigskip
\bigskip

Under the supervision of\\

\bigskip
\bigskip
\bigskip

{\bf {\Large Joseph Samuel}}\\
Theoretical Physics Department\\
Raman Research Institute, Bangalore - 560080\\

\vspace{1 cm}
\includegraphics[width=20mm]{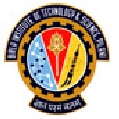}

BIRLA INSTITUTE OF TECHNOLOGY AND SCIENCE, PILANI (RAJASTHAN)\\
\bigskip

May 7th, 2010.
\end{center}

\newpage
\thispagestyle{empty}
\begin{center}
{\bf {\Large {CERTIFICATE}}}\\
\end{center}
\begin{figure}[here]
\centering
\includegraphics[scale=0.30]{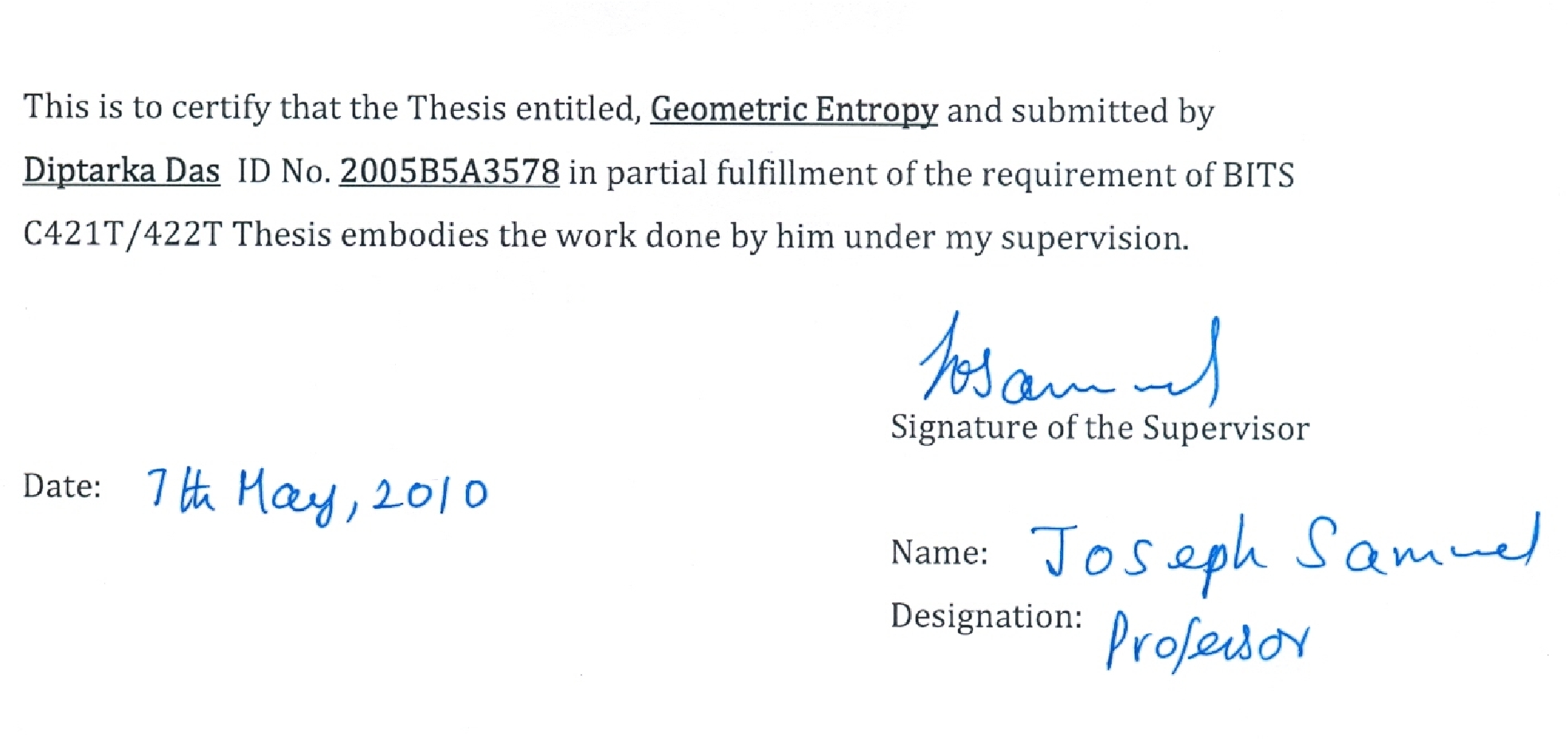}
\end{figure}
\setcounter{page}{0}

\pagestyle{empty}
\newpage
\thispagestyle{empty}
\setcounter{page}{0}
\begin{center}
\section*{\bf {\Large {ACKNOWLEDGMENTS}}}
\end{center}
\vspace{0.5 cm}
\doublespacing

I am grateful to my advisor, Sam for his guidance. He has been a great teacher, ready with lucid and very physical explanations, which helped me to assimilate the mathematical machinery in a much better way. Discussions with him on physics, not directly linked to the thesis topic has also been very productive in the development of my understanding and attitude towards physics. \\ \\
I am thankful to Dr. Alok Laddha, whose physical and mathematical insights helped me to understand the hidden subtleties. He has always been helpful and enthusiastic in discussions. Above all he has been a good friend.\\ \\ 
My sincere thanks are due to everyone who attended the talks I gave on the subject during the course of this work. They include Sam, Supurna, Alok, Arnab, Suman, Chaitra and Prasad. I also wish to thank seminar organizers at CHEP, IISc, and all those who attended my talk at IISc.\\ \\
I would like to express my thanks to BITS Pilani for providing me the opportunity to do my thesis at Raman Research Institute, Bangalore. \\ \\

-Diptarka Das

\setcounter{page}{0}

\newpage
\thispagestyle{empty}
\setcounter{page}{0}
\begin{center}
\section*{\bf \Large{ABSTRACT}}
\end{center}

The laws of mechanics of stationary black holes bear a close resemblance with the laws of thermodynamics. This is not only a mathematical analogy but also a physical one that helps us answer deep questions related to the thermodynamic properties of the black holes. It turns out that we can define an entropy which is purely geometrical for black holes. In this thesis we explain Wald's formulation which identifies black hole entropy for an arbitrary covariant theory of gravity. We would like to know precisely what inputs go into arriving at Wald's formalism. This expression for the entropy clearly depends on the precise form of the action. The secondary theme of this thesis is to distinguish thermodynamic laws which are kinematic from those which are dynamical. We would like to see explicitly in the derivation of these laws, where exactly the form of action plays a role. In the beginning we motivate the definition of entropy using the Einstein-Hilbert Lagrangian. We encounter the Zeroth law, the Hawking radiation, the second law, and then Wald's formulation.

\setcounter{page}{0}
\thispagestyle{empty}
\addtocontents{toc}{\protect\thispagestyle{empty}}
\tableofcontents
\setcounter{page}{0}
\thispagestyle{empty}

\newpage
\pagestyle{plain}
\setcounter{page}{1}

\chapter{Introduction}
Black holes are physical objects in spacetime from which nothing including light can escape. 
There is a considerable body of astronomical evidence which has emerged over the years to support 
the existence of black holes. The observations of X-ray sources reveal the presence of extremely 
energetic processes. Quasars are also known to be compact and energetic sources which astronomers 
believe to be driven by energy released during accretion onto a black hole. \\ \\
Now, given the existence of black holes, we can easily devise ways to violate the second law of thermodynamics, the second law states that the change in entropy of an isolated system can never be negative. It is easy
to think of a situation in which we take some matter with some entropy, and put it into 
the black hole. Since nothing can come out of the black hole, we conclude that the entropy of the universe has reduced, hence the change in entropy, $\de S < 0$. Therefore the second law has been violated! The way to save this ``\textit{apparent}" violation of the second law is to associate some entropy with the black hole ($S_{BH}$). This entropy, $S_{BH}$ will then increase when some matter goes into the black hole. Then we may be able to show that the net change of entropy is not negative, i.e, $\de S + \de S_{BH} \geq 0$\\ \\
It turns out that one can in fact define entropy for black holes. However the nature of this entropy is geometrical and presently we do not know if there exists any statistical description. The statistical description of the entropy may come out from some underlying quantum theory of gravity and in the classical limit it should match with the geometric one. This provides us with another motivation to study black hole thermodynamics, since it sheds some light on the nature of the quantum theory. \\ \\ 
In this thesis we study the laws of mechanics on the event horizon of a stationary black hole.\footnote{Stationary black holes are characterized by a time-independent metric.} We shall see that some of these laws are kinematic, while some depend on the dynamics of the theory. A very interesting aspect of these laws is the analogy which they bear with the laws of thermodynamics. This analogy is not only mathematical but will also help us to find physical answers to:\\ \\
$ \bullet $ \textit{Given a classical theory of gravity that admits a stationary black hole, what is the black hole entropy ?}\\
$ \bullet $ \textit{Going by the analogy with the zeroth law of thermodynamics, what characterizes equilibrium for stationary black holes ? In other words, what is the quantity that stays uniform ? Does this quantity have any relation with the temperature of the black hole ?}\\
$ \bullet $ \textit{What role does the dynamics of the theory play in these laws of mechanics ? How far can we go just by using kinematics ?}\\  \\
The rest of the chapters in this thesis try to answer the above questions as effectively as possible, building upon work done by physicists over a period of more than three decades. Chapter 2 deals with the zeroth law of black hole mechanics, which states that the quantity called ``surface gravity" is uniform over the entire event horizon of a stationary black hole. In the next chapter we encounter Hawking radiation. Following Visser,$^{\cite{Visser}}$ we derive the Hawking temperature using the bare necessities, which highlights what exactly goes in to arrive at the result. The end of chapter 3 also marks the end of how far we can reach using kinematics. The rest of the chapters make use of the equations of motion. Chapter 4 is on the second law, which for the Einstein-Hilbert Lagrangian, states that the change in the area of the event horizon of a stationary black hole is never negative. The last chapter prior to the conclusion deals with the main question, the definition of $entropy$. Here we develop a formalism due to Wald which under certain conditions will help us define entropy as a purely geometric quantity. In between, we shall apply the formalism to the Einstein-Hilbert Lagrangian and identify entropy with area. In the conclusion we go over the main results and the key points of this study. Two appendices, one on the Raychaudhuri equation and another on the Frobenius' theorem have been added for quick reference. In the rest of the introduction we familiarize ourselves with the notations followed in this thesis. We also give a list of important properties of the stationary black holes which we shall use throughout.\\ \\ 

\section{Notation and conventions}

An attempt has been made to keep the basic notation as standard as possible. Our notation follows Poisson$^{\cite{poisson}}$. The signature of the metric is assumed to be $(-1,1,1,1)$. Greek indices $(\al,\be,...)$ run from 0 to 3, latin indices $(A, B,...)$ run from 2 to 3. Geometrized units, in which $G = c = 1$, are employed. The following list of symbols will be used in this thesis: 

\begin{tabbing}
{\bf Spoken Languages} mla: \=sotiriou@sissa.it; tsotiri@phys.uoa.gr \kill
$x^\al$: \> Arbitrary coordinates on manifold $\cal{M}$ \\
$\theta^A$: \> Arbitrary coordinates on two-surface $S$ \\
$v, V$: \> Non-affine and affine parameters respectively \\
$e^\al_A = \frac{\dee x^\al}{\dee \theta^A}$: \> Holonomic basis vectors \\
$g^{\mu\nu}$: \> Lorentzian metric\\
$g$: \> Determinant of $g_{\mu\nu}$\\
$\gamma_{AB} = g_{\al \be}e^\al_A e^\be_B$: \> Induced metric on $S$ \\
$\gamma$: \> Determinant of $\gamma_{AB}$\\
$\Liek A^\al$: \> Lie derivative of $A^\al$ along $\xi^\al$\\
$\xi^\al$: \> Killing vector: $\Liek g_{\al\be} = 0$\\
$\theta, \sigma_{\al\be}, \omega_{\al\be}$: \> Expansion, shear and rotation\\
$\Gamma^{\lambda}_{\phantom{a}\mu\nu}$:  \> Affine connection\\
$\eps_{\al \be \ga \de}$:  \> Levi--Civita tensor \\
$\nabla_\mu$:  \> Covariant derivative with respect to $\{^{\lambda}_{\phantom{a}\mu\nu}\}$\\
$(\mu\nu)$:  \> Symmetrization over the indices $\mu$ and $\nu$\\
$[\mu\nu]$:  \> Anti-symmetrization over the indices $\mu$ and $\nu$\\
$R^{\lambda}_{\phantom{a}\sigma\mu\nu}$: \> Riemann tensor of $g_{\mu\nu}$\\
$R_{\mu\nu}$: \> Ricci tensor of $g_{\mu\nu}$ ($\equiv R^{\sigma}_{\phantom{a}\mu\sigma\nu}$)\\
$R$: \> Ricci scalar of $g_{\mu\nu}$ ($\equiv g^{\mu\nu}R_{\mu\nu}$)\\
$T_{\mu\nu}$: \> Stress-energy tensor $\left(\equiv -\frac{2}{\sqrt{-g}}\frac{\delta S_M}{\delta g^{\mu\nu}}\right)$\\
\end{tabbing} 
\section{Properties of a stationary black hole}
We say an asymptotically flat spacetime has a black hole if there are regions which cannot communicate with infinity, i.e., light signals sent from these do not reach infinity. The boundary between the normal region and the black hole region is called the event horizon. The event horizon is a null surface. Throughout our analysis we consider stationary black holes which have the following properties:\\
$\bullet$ The event horizon admits a Killing vector $\xi^\al = t^\al + \Omega_H \varphi^\al$, where $t^\al$ represents asymptotic time translational symmetry and $\varphi^\al$ represents asymptotic axisymmetry. \\
$\bullet$ It is null on the event horizon$^{\cite{hawk1972}}$, hence both tangent and normal to the null geodesics. It also satisfies the geodesic equation $\xi^\al_{;\be}\xi^\be = \kappa \xi^\al$. The quantity $\kappa$ which measures the failure of $\xi$ to be parallely transported is called the surface gravity. \\
$\bullet$ The null geodesics are hypersurface orthogonal. \\
$\bullet$ The null geodesics do not run into caustics on the event horizon when followed into the future.$^{\cite{penrose}}$ Once it has entered the null hypersurface, it cannot leave.\\ \\
The above properties along with some mathematical machinery will be sufficient to arrive at the laws of black hole thermodynamics and to answer the posed questions.

\chapter{Zeroth Law of Black Hole mechanics}

In this chapter we present two proofs of the zeroth law of black hole mechanics, which states that \textit{the surface gravity of a stationary black hole is uniform over the entire event horizon.}\\
The first proof is due to Bardeen, Carter and Hawking.$^{\cite{bch}}$ The proof is dependent on the Einstein-Hilbert Lagrangian and the dominant energy condition. However the zeroth law is actually kinematical, independent of the dynamics of the theory under consideration. We shall see this briefly in the last section of this chapter when we discuss the second proof due to Racz and Wald.$^{\cite{racz}}$

\section{The Bardeen, Carter and Hawking proof}\label{sec:bchproof}
The proof is two-fold. First we prove that surface gravity($\kappa$) does not change along the geodesic. Next we show that it is also uniform along the transverse directions.\\ \\ 
\textbf{Inputs to the proof}\\
\textbf{[1]} The Black hole under consideration is stationary.\\ \textbf{[2]} The above point implies, by proofs presented by Hawking and Ellis$^{\ref{lss}}$, that the event horizon is a Killing horizon.\\ \textbf{[3]} The dominant energy condition is assumed, which means that matter should follow timelike or null world lines.\\ \textbf{[4]} By the Raychaudhuri equation\footnote{See Appendix A}, stationarity also implies that matter cannot be flowing across the event horizon.\\ \textbf{[5]} The Einstein field equations.\\

Now, given $\xi^\alpha$ (tangent to the null generators on the event horizon)\footnote{Our generators are non-affinely parametrized by parameter $v$; $\xi^\al = \frac{dx^\al}{dv}$\label{nonaff}} it satisfies :
\begin{description}
\item [1] $\xi^\alpha$ is a Killing vector. 
\item [2] $\xi^\alpha$ is null on the horizon. 
\item [3] $\xi^\alpha_{;\beta}\xi^\beta = \kappa \xi^\alpha$ on horizon.\label{prop3}
\item [4] $\xi^\alpha$ has zero expansion, shear and rotation on horizon. \label{prop4}
\end{description}
Now the claim is :
\begin{equation}
\xi_{\alpha ; \beta} = (\kappa N_\alpha + c^Ae_{A\alpha})\xi _\beta - 
\xi _\alpha(\kappa N_\beta + c^Be_{B\beta})
\label{eq:Dxi}
\end{equation}
To justify equation(\ref{eq:Dxi}) we note, that if $u$ is a one-form, and $w$ is a two-form such that,
\beq
u \; \wedge\; w = 0
\label{eqw}
\eeq then we can write \beq
w = u \; \wedge\; T \quad\text{where $T$ is some other one-form.}
\label{eqw1}
\eeq 
Now by property \textbf{[4]}\ref{prop4}. $\xi^\al$ is hypersurface orthogonal (since it has vanishing rotation); so Frobenius' theorem\footnote{See Appendix B} immediately tells us : 
\beq
\xi_{[\al}\xi_{\be ;\ga]} = 0
\label{frob}
\eeq
Comparing, equations(\ref{eqw}), (\ref{eqw1}) and (\ref{frob}) we can write,
\beq
\xi_{[\be ;\ga]} = \xi_{[\be}T_{\ga]}
\label{eq:Dxi1}
\eeq
Completeness relations tell us that $N^\al$, $\xi^\al$ and $e_A^\al$ form a basis; so equation(\ref{eq:Dxi1}) can be re-written as:
\beq
\xi_{\be ;\ga} = \xi_{[\be}(aN_\ga + b^Ae_{A\ga} + c\xi_{\ga})_{]},\quad\text{where a,b and c are real coefficients}
\label{eq:Dxi2}
\eeq
But we can take $c = 0$ in the above equation, since the corresponding term after expanding out would be $c\xi_\be \xi_\ga - c\xi_\ga \xi_\be$, hence would not contribute. Thus,
\beq
\xi_{\be ;\ga} = \xi_{[\be}(aN_\ga + b^Ae_{A\ga})_{]}\label{eq:Dxi3}
\eeq
To fix $a$ in the above equation we use the geodesic equation$^{\ref{prop3}}$: \\ We compute
\beq
\begin{split}
\xi_{\al ;\be}\xi^\be \\
= \xi_{[\al}(aN_\be + b^Ae_{A\be})_{]}\xi^\be \\
= \xi_\al(aN_\be\xi^\be) = - a\xi_\al
\label{fixa}
\end{split}
\eeq
But we know $\xi_{\al ;\be}\xi^\be = \kappa\xi_\al$. Therefore from equation(\ref{fixa})
\beq
a = - \kappa
\eeq
Substituting for the value of $a$ in equation(\ref{eq:Dxi3}) we get:
\beq
\xi_{\alpha ; \beta} = (\kappa N_\alpha + c^Ae_{A\alpha})\xi _\beta - 
\xi _\alpha(\kappa N_\beta + c^Be_{B\beta})
\eeq
thereby proving our claim, i.e., equation(\ref{eq:Dxi}).\\ \\ 
Now, we would like to find the expression for $\kappa_{;\alpha}$ i.e., $\kappa_{,\alpha}$. We start with the geodesic equation for $\xi_\mu$:
\beq
\xi_{\mu ;\nu}\xi^\nu = \kappa \xi_\mu
\eeq
Differentiating both sides:
\beq
\xi_{\mu ;\nu\al}\xi^\nu + \xi_{\mu ;\nu}\xi^\nu_{;\al} = \kappa_{;\al} \xi_\mu + \kappa \xi_{\mu ;\al}
\label{zeroth11}
\eeq
We contract equation(\ref{zeroth11}) with $N^\mu$ and rearrange terms to write:
\beq
\begin{split}
\kappa_{;\al} = \kappa\xi_{\mu ;\al}N^\mu - \xi_{\mu ;\nu\al}\xi^\nu N^\mu - \xi_{\mu ;\nu}\xi^\nu_{\; ;\al}N^\mu \\
= \kappa [(\kappa N_\alpha + c^Ae_{A\alpha})\xi _\mu - \xi _\alpha(\kappa N_\mu + c^Be_{B\mu})] N^\mu \\
- R_{\mu\nu\al\be}\xi^\be \xi^\nu N^\mu - \xi_{\mu ;\nu}\xi^\nu_{\; ;\al}N^\mu\\
= \kappa (\kappa N_\al + c^Be_{B\al})- R_{\mu\nu\al\be}\xi^\be \xi^\nu N^\mu - \xi_{\mu ;\nu}\xi^\nu_{\; ;\al}N^\mu\\ 
\label{zeroth12}
\end{split}
\eeq
In the second step we used the Ricci identity for a Killing vector ($\xi_{\al; \be\ga} = R_{\al\be\ga\de}\xi^\de$) and the expansion(\ref{eq:Dxi}). In the same fashion we expand the last R.H.S term in the above equation to obtain:
\beq
\begin{split}
- \xi_{\mu ;\nu}\xi^\nu_{\; ;\al}N^\mu = (\kappa N_\nu + c^Ae_{A\nu})[(\kappa N^\nu + c^Ae_A^\nu)\xi _\al - \xi^\nu(\kappa N_\al + c^Be_{B\al})] \\
= - \kappa^2N_\al - \kappa c^Be_{B\al} - c^Ac^Be_{A\nu}e_B^\nu\xi_\al
= - \kappa^2N_\al - \kappa c^Be_{B\al} - c^Ac^B\ga_{AB}\xi_\al
\end{split}
\eeq
Putting everything back in equation(\ref{zeroth12}), we have the required expression for $\kappa_{,\al}$
\beq
\kappa_{,\al} = - R_{\mu\nu\al\be}\xi^\be \xi^\nu N^\mu - (\ga_{AB}c^Ac^B)\xi_\al
\label{eq:k,alpha}
\eeq
When equation(\ref{eq:k,alpha}) is contracted with $\xi^\al$ we immediately see that,
\beq
\kappa_{,\al}\xi^\al = - R_{\mu\nu\al\be}\xi^\be \xi^\al \xi^\nu N^\mu - (\ga_{AB}c^Ac^B)\xi_\al\xi^\al = 0
\label{eq:zeroth1}
\eeq
In the second step, the first term is zero since $R_{\mu\nu\al\be}$ is antisymmetric in ($\al$,$\be$) and the term $\xi^\be\xi^\al$ is symmetric in ($\al$,$\be$). The second term is zero since $\xi^\al$ is a null vector. Equation(\ref{eq:zeroth1}) proves that surface gravity does not change along the geodesic.
\newpage
Now to show that the surface gravity $\kappa$ is constant over the entire horizon we need to show that $\kappa_{,\alpha}e^\alpha_A = 0$, along with equation(\ref{eq:zeroth1}). So we evaluate it next. 
Using equation(\ref{eq:k,alpha}) we get,
\begin{equation}
 \kappa_{,\alpha}e^\alpha_A = -R_{\alpha\beta\gamma\delta}N^\gamma\xi^\delta\xi^\beta e_A^\alpha 
\label{eq:zeroth21}
\end{equation} since $\xi_\al e^\al_A = 0$.
Now we use the completeness relation:
\begin{equation}
 g^{\beta\gamma} = -\xi^\gamma N^\beta - N^\gamma \xi^\beta + \ga^{BC}e_B^\beta e_C^\gamma
\end{equation}
to re-express equation(\ref{eq:zeroth21}) as:
\begin{equation}
 \kappa_{,\alpha}e^\alpha_A = R_{\alpha\beta\gamma\delta}\xi^\delta e_A^\alpha(\xi^\gamma N^\beta - \ga^{BC}e_B^\beta e_C^\gamma + g^{\beta\gamma})
\end{equation} Expanding the terms,
\begin{equation}
 \kappa_{,\alpha}e^\alpha_A = R_{\alpha\beta\gamma\delta}\xi^\delta\xi^\gamma e^\alpha_A N^\beta + g^{\beta\gamma}R_{\alpha\beta\gamma\delta}\xi^\delta e_A^\alpha - \ga^{BC}R_{\alpha\beta\gamma\delta} e_A^\alpha e_B^\beta e_C^\gamma\xi^\delta
\end{equation}
The first term on the right hand side of the above equation vanishes since $R_{\mu\nu\alpha\beta}$ is antisymmetric in ($\alpha$,$\beta$) and the term $\xi^\alpha\xi^\beta$ is symmetric in ($\alpha$,$\beta$). We get:
\begin{equation}
 \kappa_{,\alpha}e^\alpha_A = g^{\beta\gamma}R_{\alpha\beta\gamma\delta}\xi^\delta e_A^\alpha - \ga^{BC}R_{\alpha\beta\gamma\delta} e_A^\alpha e_B^\beta e_C^\gamma\xi^\delta
\end{equation}
or,\begin{equation}
 \kappa_{,\alpha}e^\alpha_A = -R_{\alpha\beta}e_A^\alpha\xi^\beta - \ga^{BC}R_{\alpha\beta\gamma\delta} e_A^\alpha e_B^\beta e_C^\gamma\xi^\delta
\label{eq:zeroth22}
\end{equation}
Now we consider the quantity: $B_{AB} = \xi_{\alpha;\beta}e^\alpha_Ae^\beta_B$ If we use equation(\ref{eq:Dxi}) to expand $\xi_{\alpha;\beta}$ then we end up with: 
\beq
B_{AB} = [(\kappa N_\alpha + c^Ae_{A\alpha})\xi _\beta - 
\xi _\alpha(\kappa N_\beta + c^Be_{B\beta})]e^\alpha_Ae^\beta_B\\
\eeq
Therefore,\begin{equation}
 B_{AB} = 0
\end{equation}
We also note that the tangential derivatives of $B_{AB}$ vanish on the horizon. It implies: 
\begin{equation}
 (B_{AB})_{;\gamma}e^\gamma_C = 0
\end{equation}
or, \begin{equation}
 \xi_{\alpha;\beta\gamma}e^\alpha_Ae^\beta_B e^\gamma_C = 0
\end{equation}
or, \begin{equation}
 R_{\alpha\beta\gamma\delta} e_A^\alpha e_B^\beta e_C^\gamma\xi^\delta = 0
\label{eq:vanish}
\end{equation}
So, using the above equation, equation(\ref{eq:zeroth22}) simplifies to:
\begin{equation}
 \kappa_{,\alpha}e^\alpha_A = -R_{\alpha\beta}e_A^\alpha\xi^\beta
\label{eq:zeroth23}
\end{equation}
Now we invoke the Einstein equation:
\begin{equation}
 R_{\alpha\beta} - \frac{1}{2}Rg_{\alpha\beta} = 8\pi T_{\alpha\beta}
\label{eq:einstein}
\end{equation}
Multiplying the equation by $e_A^\alpha\xi^\beta$ and carrying out the sum over $\alpha$ and $\beta$ we get:
\begin{equation}
 R_{\alpha\beta}e_A^\alpha\xi^\beta - \frac{1}{2}Rg_{\alpha\beta}e_A^\alpha\xi^\beta = 8\pi T_{\alpha\beta}e_A^\alpha\xi^\beta
\end{equation}
But $g_{\alpha\beta}e_A^\alpha\xi^\beta = 0$ so we get:
\begin{equation}
 R_{\alpha\beta}e_A^\alpha\xi^\beta = 8\pi T_{\alpha\beta}e_A^\alpha\xi^\beta
\end{equation}
Substituting this result into equation(\ref{eq:zeroth23}) we obtain:
\begin{equation}
 \kappa_{,\alpha}e^\alpha_A = 8\pi j_{\alpha}e_A^\alpha
\label{eq:zeroth24}
\end{equation} where $j_\alpha = -T_{\alpha\beta}\xi^\beta$.\\ 
According to the assumption of the dominant energy condition, $j_\alpha$ which represents the flux being carried away, must be timelike or null. Which implies:
\begin{equation}
 j_\alpha j^\alpha \leq 0
\label{ineq:cond1}
\end{equation}
By the stationarity assumption, we know\bea \frac{d\theta}{d\tau} = 0\\\text{and,\qquad} \theta = 0 \eea Using the Raychaudhuri and the Einstein's equations this gives:
\begin{equation}
 T_{\alpha\beta}\xi^\alpha\xi^\beta = 0
\label{eq:cond2}
\end{equation}
Note, that $\xi^\alpha$, $N_\alpha$ and $e_{A\alpha}$ forms a complete basis, so we can write:
\begin{equation}
 j_\alpha = A\xi_\alpha + BN_\alpha + C^Ae_{A_\alpha}
\label{eq:j}
\end{equation} where A,B and C are real numbers.
Consider now, $j_\alpha\xi^\alpha$, the above equation gives, 
\begin{equation}
 j_\alpha\xi^\alpha = -B
\end{equation}
And using the definition of $j_\alpha$ and equation(\ref{eq:cond2}) we get $j_\alpha\xi^\alpha = 0$. Therefore $B = 0$. So equation(\ref{eq:j}) is simplified to:
\begin{equation}
 j_\alpha = A\xi_\alpha + C^Ae_{A_\alpha}
\label{eq:j1}
\end{equation}
Now consider, $j_\alpha j^\alpha$.
\begin{equation}
 j_\alpha j^\alpha = (A\xi_\alpha + C^Ae_{A_\alpha})(A\xi_\alpha + C^Be_{B_\alpha})
\end{equation}
\begin{equation}
 = C^2
\end{equation}
Now inequality(\ref{ineq:cond1}) says that $C^2 \leq 0$ but we have assumed that C is a real number. Hence the only possibility is $C = 0$. So we are left with:
\begin{equation}
 j_\alpha = A\xi_\alpha
\end{equation}
So $j_\alpha$ is parallel to $\xi_\alpha$. Therefore clearly,
\begin{equation}
 j_{\alpha}e_A^\alpha = 0
\end{equation}
Hence equation(\ref{eq:zeroth24}) gives us:
\begin{equation}
 \kappa_{,\alpha}e^\alpha_A = 0
\label{eq:zeroth2}
\end{equation}
Equations (\ref{eq:zeroth1}) and (\ref{eq:zeroth2}) are all we needed to show that surface gravity, $\kappa$ is constant over the entire event horizon.

\section{The Zeroth Law is \textit{kinematical}}\label{sec:zerothwald}
We notice that in the previous section the second part of the proof depended on the dynamics of the theory. However it was possible to show that $\kappa$ is constant along a geodesic without the use of Einstein's equations.(see equation(\ref{eq:zeroth1}) Kinematically we were able to go as far as equation(\ref{eq:zeroth23}),
\beq
\kappa_{,\alpha}e^\alpha_A = -R_{\alpha\beta}e_A^\alpha\xi^\beta
\eeq
Using the arguments in Racz and Wald's paper$^{\cite{racz}}$ we would like to show that the R.H.S of the above equation is zero kinematically.\\ \\
We assume that the event horizon is geodesically complete (in the sense that the generators never leave the horizon when we go back into the past) and surface gravity, $\kappa$, is non-zero. We can reparametrize our generators affinely.\footnote{Our generators are non-affinely parametrized by parameter $v$; $\xi^\al = \frac{dx^\al}{dv}$\label{nonaff}} Let us call the affine parameter $V$\footnote{We call the affinely parametrized generators as $k^\al$}, then it is related to the non-affine parameter $v$ through:
\beq
\frac{dV}{dv} = e^{\kappa v} 
\eeq
or,
\beq
V = \frac{e^{\kappa v}}{\kappa} \label{affine}
\eeq
Therefore,
\bea
\xi^\al = \frac{dx^\al}{dv} = \frac{dx^\al}{dV}\frac{dV}{dv} \\
= k^\al e^{\kappa v}
\eea
Using equation(\ref{affine})
\beq
\xi^\al = \kappa V k^\al \label{zeroth3}
\eeq
From equation(\ref{zeroth3}), we see that $\xi^\al\rightarrow 0$ as $V\rightarrow 0$. Since the horizon is geodesically complete, as we go back into the past, $\xi^\al$ is zero at a point. This defines the \textit{bifurcation two-sphere}.\label{bifurcate}\\
Thus, existence of \textit{bifurcation two-sphere} implies that R.H.S of equation(\ref{eq:zeroth23}) vanishes. So, $\kappa$ stays the same as we go from generator to generator. Therefore we conclude that the surface gravity stays uniform over the event horizon of a stationary black hole. It can be seen that the zeroth law holds irrespective of the existence of a bifurcate horizon. Consider two black holes which are identical at some finite,$v > 0$. Let one of them be an eternal black hole for which equation(\ref{eq:zeroth2}) holds. Since the spacetime is assumed to be continuous the same must be true for the second black hole as well. Therefore surface gravity is constant over its entire event horizon.

\chapter{Surface gravity is temperature}
In this chapter we shall encounter an important result relating surface gravity to temperature. This result is due to Hawking$^{\cite{Hawking}}$. We shall derive this result using the minimalistic approach as followed by Visser$^{\cite{Visser}}$. The analysis is purely kinematical. The Einstein's equations are not used anywhere. We shall look at the generic features of the modes near the horizon using the eikonal approximation. Specifically we shall look for a Boltzmann factor. We assume a spherically symmetric metric with a horizon. \\ \\
\section{Surface gravity in terms of the \Painleve-Gullstrand coordinates}
In general relativity any spherically symmetric geometry can be put into the following form:
\beq
ds^2 = - [c(r,t)^2 - v(r,t)^2]\; dt^2 - 2 v(r,t) dr \; dt  
+ dr^2 + r^2 [d\theta^2 + \sin^2\theta \; d\phi^2 ]
\eeq
The metric is called \Painleve-Gullstrand which in matrix form looks like:
\beq
g_{\mu\nu}(t,\vec x) 
\equiv
\left[ {\begin{array}{cc}
 -(c^2 - v^2) & -v \hat r_j  \\
 -v \hat r_i & \delta_{ij}  
 \end{array} } \right]
\label{pg}
\eeq 	
The apparent horizon is located at $c(r,t) = |v(r,t)|$, and for stationary black holes this matches with the event horizon. Now we define:
\begin{equation}
g_H(t) 
= {1\over2} \left.{\d  [c(r,t)^2 - v(r,t)^2]\over \d r}\right|_H 
= c_H \left.{\d  [c(r,t)- |v(r,t)|]\over \d r}\right|_H
\end{equation}
and\begin{equation}
\kappa = {g_H\over c_H}
\label{sg1}
\end{equation}
If the geometry is stationary, this reduces to the ordinary definition of
{surface gravity}. In Chapter 1, our defining equation for surface gravity, $\kappa$, was:
\beq
\xi_{;\be}^\al \xi^\be  = \kappa \xi^\al
\eeq
We can see that for a timelike Killing vector $\xi^\al_{(t)} = (1, \vec{0})$ the above equation reduces to:
\beq
\Gamma^0_{00}\xi^0 = \kappa \xi^0
\label{sg2}
\eeq
When the given metric(\ref{pg}) is stationary, it is easy to check that,
\beq
\Gamma^0_{00} = \left.{\d  [c(r,t)- |v(r,t)|]\over \d r}\right|_H
\label{sg3}
\eeq
Comparison of equations (\ref{sg1}), (\ref{sg2}) and (\ref{sg3}) shows that the two definitions of surface gravity match for the static case. If the geometry is not stationary, then equation(\ref{sg1}) is taken as the definition of ``surface gravity''.$^{\cite{Visser}}$

\section{Eikonal approximation ($s$ wave)}
We consider a scalar quantum field $\phi(r,t)$ on this \Painleve-Gullstrand background and take the eikonal approximation for the $s$ wave.
\begin{equation}
\phi(r,t)  
= {\cal A}(r,t) \; \exp[\mp i\varphi(r,t)] =   {\cal A}(r,t) \;
{
\exp\left[\mp i\left(\omega\;t - \int^r k(r') \; \d r'\right)\right]}
\label{phi}
\end{equation}
where the field is written as a rapidly varying phase times a slowly varying envelope. With the Lagrangian density, $\frac{1}{2}\dee_\mu \phi \dee^\mu \phi$, the equation of motion becomes
\beq
\Box \phi = 0
\eeq
or,\beq
\begin{split}
\Box{\cal A}\exp[\mp i\varphi] \mp i\dee_\mu{\cal A}\dee^\mu \ph \exp[\mp i\varphi] \mp i\Box \ph{\cal A}\exp[\mp i\varphi]\\
\mp i\dee_\mu \ph \dee^\mu{\cal A}\exp[\mp i\varphi] + \dee^\mu \ph \dee_\mu \ph{\cal A}\exp[\mp i\varphi] = 0
\end{split}
\eeq
In the eikonal approximation only the last term survives. We use Feynman's ``$i\epsilon$-prescription'' ($\epsilon$ is real, positive, and infinitesimal). The wave equation reads:
\begin{equation}
g^{\mu\nu} \; \partial_\mu \varphi \; \partial_\nu \varphi +i\epsilon = 0.
\label{main}
\end{equation}
Note that in invoking the prescription we have used the fact that the spacetime geometry is smooth, even at the horizon. Putting in the metric(\ref{pg}) in equation(\ref{main}) we obtain,
\begin{equation}
\omega - v k = \sigma\; (1+i\epsilon) \; c k; \qquad \sigma = \pm1.
\end{equation}
Solving for $k(r,t)$, (for specific real frequency $\omega$) we have:
\begin{equation}
k 
= 
{\omega\over  \sigma\; (1+i\epsilon) \; c + v} 
=
{\sigma \; \omega \over  (1+i\epsilon) \; c + \sigma v} 
= {\sigma \;  (1+i\epsilon) \; c - v\over  (1+i\epsilon)^2\; c^2-v^2} \;\omega.
\label{krt}
\end{equation}
Note:
\begin{equation}
\sigma = +1 \qquad \implies \qquad \hbox{outgoing mode}
\end{equation}
\begin{equation}
\sigma = -1 \qquad \implies \qquad \hbox{ingoing mode}
\end{equation}
Now, it turns out that one can estimate the functional form of ${\cal A}$ from current conservation arguments. The current approximately is,
\begin{equation}
J_\mu = | {\cal A}(r,t) |^2 \;\; ( \omega, k, 0,0 )
\end{equation}
Then
\begin{equation}
\nabla_\mu \, J^\mu = 0 \qquad \implies \qquad\qquad
| {\cal A}(r,t) | \propto {1\over r}
\label{Art}
\end{equation}
So using equations(\ref{Art}) and (\ref{krt}) we can write down the quantum field $\phi^{\ref{phi}}$ as,
\begin{equation}
\phi(r,t)  \approx {{\cal N}\over r}\;
\exp\left[\mp i\left(\omega\;t - \int^r  {\omega\over  \sigma\; (1+i\epsilon) \; c(r') + v(r')} \; \d r'\right)\right],
\label{mainwave}
\end{equation}
where ${\cal N}$ is some normalization.
%%%%%%%%%%%%%%%%%%%%%%%%%%%%%%%%%%%%%%%%%%%%%%%%%%%%%%%%%%%%%%%%%%%%%%
\section{Outgoing and straddling modes}
We consider the outgoing mode $\sigma = +1$
\begin{equation}
k_{\mathrm{out}} 
= 
{\omega\over  (1+i\epsilon) \; c + v}
\end{equation}
In the vicinity of the future horizon $r\approx r_H$ (with $v
\approx - c$) the outgoing wavevector is:
\begin{equation}
k_{\mathrm{out}} 
\approx 
{\omega \over [g_H/c_H] (r-r_H) + i \epsilon \; c_H}
\end{equation}
Rewriting this in terms of the ``principal part" and a delta function contribution we have,
\begin{equation}
k_{\mathrm{out}} 
\approx
{c_H \;\omega\over g_H} 
\left\{ \wp\left({1\over r-r_H}\right) - i \pi \; \delta(r-r_H)\right\}.
\end{equation}
Since we are not crossing the horizon in this case, we can ignore the $i\epsilon$. Therefore just outside the horizon we have:
\begin{equation}
\int^r k = \int^r {dr' \;\omega\over c(r') - |v(r')|} \approx 
\int^r {dr'\; c_H \;\omega\over g_H (r' - r_H)} 
= { {c_H \;\omega\over g_H} \; \ln[r- r_H]}
\end{equation}
Thus the field for $r>r_H$ takes the following form,
\begin{eqnarray*}
\phi(r,t)_{\mathrm{out}} &\approx& {\cal N}_{\mathrm{out}}\;
{\exp\left(\pm i \left[\omega c_H\over g_H\right] \;  \ln[r- r_H]\right)
\over r_H} 
\;\exp\left\{
\mp i \omega t
\right\}
\\
&\approx& 
 {\cal N}_{\mathrm{out}}\;
{{[r- r_H]^{\pm i\omega c_H/g_H}}\over r_H} 
\;\exp\left\{
\mp i \omega t
\right\}
\end{eqnarray*}
In terms of surface gravity, $\kappa$, we have;
\beq
\phi(r,t)_{\mathrm{out}} \approx 
 {\cal N}_{\mathrm{out}}\;
{{[r- r_H]^{\pm i\omega / \kappa}}\over r_H} 
\;\exp\left\{
\mp i \omega t
\right\}
\label{outmode}
\eeq
Now we look at those ``outgoing" modes which straddle the horizon. In this case we cannot ignore the $i\epsilon$ contribution while calculating the wavevector, since we are crossing the horizon. So now our wavevector is:
\begin{eqnarray}
\int^{r^+}_{r_-} k_{\mathrm{out}} &\approx&
\int^{r^+}_{r_-} \d r' \; {c_H \; \omega \over g_H} 
\left\{ \wp\left({1\over r'-r_H}\right) - i \pi \; \delta(r'-r_H)\right\} 
\nonumber\\
&=&
{c_H\; \omega\over g_H} \; 
\left\{ \ln{|r_+ - r_H|\over|r_- - r_H|} - i \pi\right\}.
\end{eqnarray}
On putting this back into equation(\ref{mainwave}) we can write the straddling field in terms of the Heaviside function as:
\begin{eqnarray}
\phi(r,t)_{\mathrm{straddle}} &\approx& 
{\cal N}_{\mathrm{straddle}}\;
\left[ 
\Theta(r_H-r) \; {\exp\left\{ +{\pi\;\omega \;c_H\over g_H}\right\}} +
\Theta(r-r_H)
\right]
\nonumber\\
&& 
\times {{|r- r_H|^{\pm i\omega c_H/g_H}}\over r_H} \; 
\exp\left[
\mp i \omega t
\right]
\end{eqnarray}
We see that this mode picks up an exponential factor which contains the surface gravity
\beq
\exp\left\{ +{\pi\omega\over\kappa}\right\}
\eeq
This clearly indicates a relation between temperature and surface gravity.\footnote{Occurence of such factors were key to Hawking's derivation$^{\cite{Hawking}}$.} We shall investigate this relationship more carefully by matching the current at the horizon. We have,
\[
|{\cal N}_{\mathrm{straddle}}|^2 
\left[ {\exp\left\{ +{2\pi\;\omega\over \kappa}\right\}} - 1 \right]
=
|{\cal N}_{\mathrm{out}}|^2
\]
The ratio of the normalizations is nothing but the Planckian distribution for the outgoing flux. We see that,
\beq
\left|{ {\cal N}_{\mathrm{straddle}}\over {\cal N}_{\mathrm{out}}} \right|^2  =
{1\over {\exp\left\{ +{2\pi\;\omega\over\kappa}\right\}} - 1 }.
\label{hawk}
\eeq 
At this point, it is useful to see the analogy by treating the black hole as a black body. Any black body which absorbs in presence of radiation, also radiates. In addition to stimulated emission of radiation, there must be spontaneous emission. Einstein's work $^{\cite{eins}}$showed that the rate coefficients of stimulated emission and absorption are the same. And the ratio of spontaneous emission to stimulated emission is given by the Planck distribution law, at the black body equilibrium temperature $T$.\\
The same argument has been drawn upon by Hawking in his original treatment$^{\cite{Hawking}}$. Using QFT(for bosonic fields) in curved spacetime he was able to show, that the total number of particles created and emitted to infinity is,
\beq
{1\over {\exp\left\{ +{2\pi\;\omega\over\kappa}\right\}} - 1 }\Gamma
\eeq
where, $\Gamma$ is the fraction of the wavepacket which will enter the black hole and $\kappa$ is the surface gravity. We have obtained the identical ratio in equation(\ref{hawk}). It is the relation between spontaneous emission and absorption coefficients. Thus we can readily read out the ``Hawking" temperature as,
\begin{equation}
T_H = {\hbar \over 2\pi k}\kappa
\label{kappaisT}
\end{equation}
Based on the above derivation we conclude that surface gravity is in fact Hawking temperature. It is to be noted that the above derivation is completely kinematical, making no use of the Einstein's equations describing the dynamics of the gravitational field. Hence independent of the Lagrangian we can relate temperature to surface gravity.

\chapter{The Second Law of Black Hole mechanics}
We have seen in the last two chapters some kinematical properties of the event horizon of a stationary black hole. In this chapter we analyze further the geometry of the event horizon using the tools that we have, namely the Raychaudhuri equation \footnote{Appendix A.\label{rcfoot}} and the Frobenius' theorem. \footnote{Appendix B.} We will try and see if the geometry tells us anything more. The treatment in this chapter is not completely kinematical as we shall impose certain restrictions on the Lagrangian.
\section{Focusing theorem}
\begin{figure}[!h]
    \begin{center}
        \scalebox{0.45}{
            \includegraphics[scale=.75]{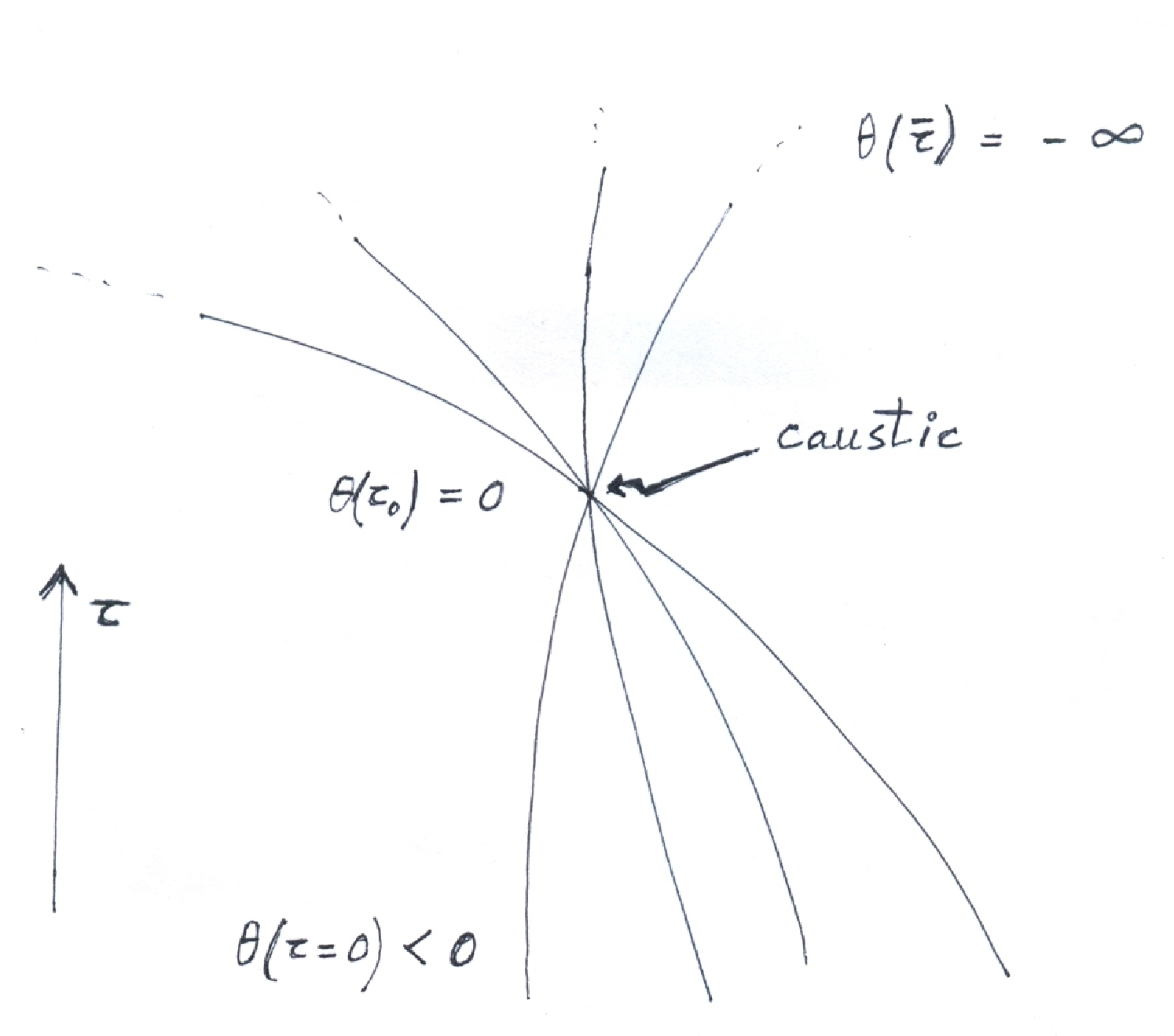}}
    \end{center}
            \caption{\small Evolution of the congruence according to equation(\ref{conth}) when $\theta(\tau = 0) < 0$}
            \label{caust}
\end{figure}

Let a congruence of null geodesics be \textit{hypersurface orthogonal}.  Then Frobenius' theorem tells us that the rotation tensor $\omega_{\al\be} = 0$. The Raychaudhuri$^{\ref{rcfoot}}$ equation then implies
\beq
\frac{d\theta}{d\tau} = - \frac{1}{2}\theta^2 - \sigma^2 - R_{\al\be}k^\al k^\be
\label{rcnull1}
\eeq
The first two terms on the right hand side give a negative contribution. The third term is a dynamical one since the Ricci tensor $R_{\al\be}$ is related to the energy-momentum tensor through the field equations.\\
%We will now restrict ourselves to the case where
%\beq
%R_{\al\be}k^\al k^\be \geq - \frac{1}{2}\theta^2 - \sigma^2
%\label{conR}
%\eeq
%In such a case we notice that, 
%\beq
%\frac{d\theta}{d\tau} \leq 0
%\eeq
%The above equation means that the expansion must decrease during the congruence's evolution. Thus initially diverging congruence will diverge less rapidly and initially converging ones will converge more rapidly. In case of the Einstein-Hilbert Lagrangian (where equation(\ref{conR}) is satisfied for \textit{null energy condition}\footnote{$T_{\al\be}k^\al k^\be \geq 0$, where $T_{\al\be}$ is the energy-momentum tensor}) the physical interpretation is that gravitation  is an attractive force when the null energy condition holds. And it is due to this attraction that the geodesics get focused. This is the statement of the focusing theorem.\\
Now we make a restriction on equation(\ref{rcnull1}) by demanding:
\beq
R_{\al\be}k^\al k^\be \geq 0 
\label{conR1}
\eeq
The consequence of the above inequality on the Raychaudhuri equation is:
\beq
\frac{d\theta}{d\tau} \leq - \frac{1}{2}\theta^2 
\label{rcnull2}
\eeq
Integrating equation(\ref{rcnull2}) yields,
\beq
\theta^{-1}(\tau) \geq \theta^{-1}(\tau = 0) + \frac{\tau}{2}
\label{conth}
\eeq
This shows that if the congruence is initially convergent i.e, $\theta(\tau = 0) < 0$ then within \bea \tau \leq \frac{2}{|\theta(\tau = 0)|} \\\text{we have, \;\;} \theta(\tau)\rightarrow - \infty\label{focus}\eea The interpretation of this result is that the congruences which were converging have focused and then diverged. Therefore they must have developed a $caustic$,[see Figure(\ref{caust})] a point at which some of the geodesics come together. You can understand the use of the term ``caustic'' by focusing sun rays onto the palm of your hand using a magnifying glass.
\section{Interpretation of $\theta$ and the Area theorem}
For the null case, we now prove : 
\beq
\theta = \frac{1}{\sqrt{\ga}}\frac{d}{d\tau}(\sqrt{\ga})\label{thetasl}
\eeq 
where, $\ga = \det(\ga_{AB})$. Here, $\ga_{AB}$ is the 2-dimensional metric on the cross section of the null congruence satisfying:
\beq
\ga^{\al\be} = \ga^{AB}e^\al_Ae^\be_B \label{gab1}
\eeq
and,
\beq
\ga_{AB} = \ga_{\al\be}e^\al_Ae^\be_B \label{gab2}
\eeq
Starting from the R.H.S of equation(\ref{thetasl}) we have,
\beq
\frac{1}{\sqrt{\ga}}\frac{d}{d\tau}(\sqrt{\ga}) = \frac{1}{2}\ga^{AB}\frac{d}{d\tau}(\ga_{AB})\label{thetasl2}
\eeq
Using equation(\ref{gab2}) and the completeness relation\footnote{$\ga_{\al\be} = g_{\al\be} + u_\al N_\be + u_\be N_\al$} we have:
\bea
\frac{d}{d\tau}(\ga_{AB}) = (g_{\al\be}e_A^\al e^\be_B)_{;\mu}k^\mu \\
= g_{\al\be}[(e_{A;\mu}^\al k^\mu)e_B^\be + (e_{B;\mu}^\be k^\mu)e_A^\al]\\
= g_{\al\be}[k^\al_{;\mu}e_A^\mu e_B^\be + k^\be_{;\mu}e_B^\mu e_A^\al]\\
= (B_{\al\be} + B_{\be\al})e_A^\al e_B^\be
\eea
The third step follows from the fact that, $\mathcal{L}_{\bf{k}}e_A^\al =0$ and the fourth step follows from the definition of the tensor $B_{\al\be}$.\footnote{See Appendix A.} Going back to equation(\ref{thetasl2}) we have,
\beq
\frac{1}{\sqrt{\ga}}\frac{d}{d\tau}(\sqrt{\ga}) = \frac{1}{2}\ga^{AB}(B_{\al\be} + B_{\be\al})e_A^\al e_B^\be
\eeq
Using equation(\ref{gab1}),
\beq
\frac{1}{\sqrt{\ga}}\frac{d}{d\tau}(\sqrt{\ga}) = \ga^{\al\be}B_{\al\be} = \theta
\eeq
The above result proves equation(\ref{thetasl}). Now we consider the congruence's area element, 
\beq \de A = \oint \sqrt{\ga} d^2x 
\eeq 
Rate of change of this quantity with respect to the affine parameter is: 
\beq
\frac{d}{d\tau}\de A = \oint \frac{d}{d\tau}(\sqrt{\ga}) d^2x 
\eeq
Using equation(\ref{thetasl}) we can rewrite the above equation as:
\beq
\frac{d}{d\tau}\de A = \oint \sqrt{\ga}\theta d^2x 
\eeq
or,\beq
\theta = \frac{1}{\de A}\frac{d}{d\tau}\de A
\label{sltheta}
\eeq
Thus $\theta$ is the fractional rate of change of the congruence's cross-sectional area. The area theorem follows directly from the Focusing theorem and equation(\ref{sltheta}).\\ \\  It was observed by Penrose that the event horizon is generated by null geodesics with no future end points. This means that the null generators cannot run into caustics. Therefore the focusing theorem implies that the expansion parameter, $\theta$, has to be either positive or zero, everywhere on the event horizon. This is true because, in case $\theta$ was negative, then by equation(\ref{focus}), we would have a caustic. Therefore everywhere on the event horizon, $\theta \geq 0$.From equation(\ref{sltheta}) we thus have the result, that the event horizon area will not decrease, i.e,
\beq
\de A \geq 0
\label{areath}
\eeq
This is the area theorem, or the second law of black hole mechanics.

\chapter{Wald's formulation and the first law}

We consider a general, classical theory of gravity in \textit{n} dimensions, arising from a diffeomorphism invariant Lagrangian. In any such theory, to each vector field, $\xi^\al$, on spacetime one can associate a local symmetry and hence, a Noether current, (\textit{n}-1)-form, \bJ, and (on-shell) a Noether charge (\textit{n}-2)-form, \bQ, both of which are locally constructed from $\xi^\al$ and the fields appearing in the Lagrangian. \cite{w1} 

Using the Noether current we shall derive the first law of black hole mechanics for stationary black holes with a bifurcate Killing horizon (so that the Zeroth law holds).\cite{w2} This is the "physical process version" of the first law, in which we pass from an initial stationary state to a final stationary state. In between the black hole interacts with its environment exchanging energy and angular momentum and is \textit{not} stationary. From the first law we can then conclude that the black hole entropy is simply surface integral of the (\textit{n}-2)-form Noether charge associated with the horizon Killing field. \cite{w1}
\clearpage
\newpage

\section{Constraints on the variation of the Hamiltonian of a diffeomorphism invariant theory}

We shall now formulate the Hamiltonian for a diffeomorphism invariant Lagrangian in \textit{n}-dimensional manifold. We shall view the Lagrangian as an \textit{n}-form, \bL, rather than as a scalar density. At each point in the spacetime, $\bL$ is required to be a function of the spacetime metric $g_{\al \be}$, as well as other matter fields and finitely many of its derivatives at the point. The higher derivative theories of gravity are included in this framework.\\We use the symbol '$\phi$' to denote all the dynamical fields, including the metric. Now diffeomorphism invariant theories mean, for any diffeomorphism,\\ $\psi$ : $M \xrightarrow{} M  $, we have,
\beq
\bL[\psi^*(\phi)] = \psi^*\bL[\phi]
\eeq
The above equation says that if we pull back $\phi$ and then evaluate \bL, or we evaluate $\bL$ and then do the pull back, we should end up with the same result. Now we carry out a first order variation of \bL. This can be written as (see \cite{w5} ):
\beq
\de\bL = \bE\de\phi + d\bTh
\label{de-l}
\eeq
The Euler-Lagrange equations of motion of the theory are simply $\bE = 0$. The above equation defines $\bTh$. The (\textit{n}-1)-form, $\bTh$, is locally constructed from $\phi$ and $\de\phi$, and we can use it to define the symplectic current $\Om$ as:
\beq
\Om(\phi,\de_1\phi,\de_2\phi) = \de_1[\bTh(\phi,\de_2\phi)] - \de_2[\bTh(\phi,\de_1\phi)]
\label{defOm}
\eeq
Now, let $\xi^\al$ be any vector field on $M$. Consider the field variation $\de\phi = \Liek\phi$. The diffeomorphism invariance of $\bL$ implies that under this variation, 
\beq
\de\bL = \Liek\bL
\label{deL}
\eeq
Using the following identity: 
\beq
\Liek\Lambda = \xi\cd d\Lambda + d(\xi\cd\Lambda)
\label{lie}
\eeq
We can rewrite equation(\ref{deL}) as:
\beq
\de\bL = \Liek\bL = d(\xi\cd\bL)
\label{del}
\eeq
since $d\bL$ vanishes, as it is an $(n+1)$-form in an $n$ dimensional space.\\
Now, equation(\ref{del}) indicates that the Lagrangian changes by a total derivative. Hence we can associate a Noether current (\textit{n}-1)-form, $\bJ$ to each $\xi^\al$, defined by:
\beq
\bJ = \bTh(\phi,\Liek\phi) - \xi\cd\bL
\label{defJ1}
\eeq
We can check, 
\beq
\beg
d\bJ = d\bTh - d(\xi\cd\bL)\\
d\bJ = \de\bL - \bE\de\phi - \de\bL\\
d\bJ = -\bE\de\phi
\eeg
\eeq
so, $\bJ$ is closed whenever the equations of motion are satisfied($\bE = 0$). One can then show, (see the Appendix of \cite{w3}) that $\bJ$ can always be written in the form:
\beq
\bJ = d\bQ + \xi^\al \bK_\al
\label{defJ2}
\eeq
where, $\bQ$ is the Noether charge (\textit{n}-2)-form, and $\bK_\al$ are the constraints of the theory.\\
Going back to the original definition of $\bJ$, equation(\ref{defJ1}) the first order variation in $\bJ$ due to arbitrary variation of $\de\phi$ is,
\beq
\de\bJ = \de\bTh - \xi\cd\de\bL
\eeq
Now, using equation(\ref{de-l}) we write the above equation as:
\beq
\de\bJ = \de\bTh - \xi\cd[\bE\de\phi + d\bTh]
\eeq
Using equation(\ref{lie}):
\beq
\de\bJ = \de[\bTh(\phi,\Liek\phi)] - \Liek[\bTh(\phi,\de\phi)] + d(\xi\cd\bTh)
\eeq
Now, by the definition of the symplectic current $\Om$ (equation(\ref{defOm})), the first two terms can be combined, and the above equation can be re-written as:
\beq
\de\bJ = \Om(\phi,\de\phi,\Liek\phi) + d(\xi\cd\bTh)
\label{delj}
\eeq
Thus the current $\Om$ is:
\beq
\Om(\phi,\de\phi,\Liek\phi) = \de\bJ - d(\xi\cd\bTh)
\label{Om}
\eeq
When the above equation is integrated over a Cauchy surface (\textit{slice}), $\Sigma$, comparison with the Hamilton's equations of motion shows that if a Hamiltonian, $H$, exists then it must satisfy\cite{w6}\cite{w2},
\beq
\de H = \de \int_\Sigma\bJ - \int_\Sigma d(\xi\cd\bTh)
\label{deH1}
\eeq
\clearpage
\newpage

\section{The First Law of Black Hole Mechanics}
\begin{figure}[!h]
    \begin{center}
        \scalebox{0.45}{
            \includegraphics[scale=.75]{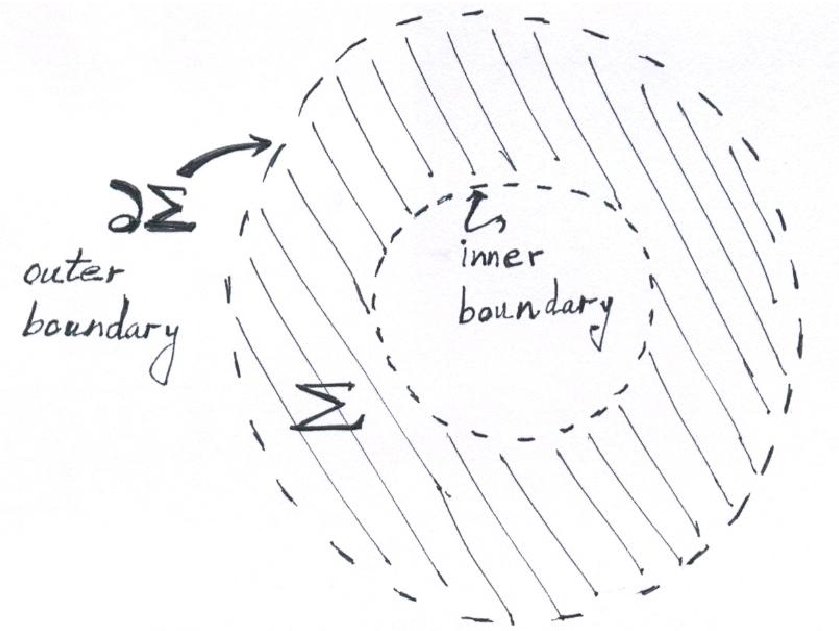}}
    \end{center}
            \caption{\small \textit{Slice} boundary approaches $\infty$, so $\de\bK_\al = 0$}
            \label{surface}
\end{figure}

In this section we shall derive the physical process version of the first law of Black Hole mechanics. We start by deriving the formulae for first-order variations in ADM mass and angular momentum (using the Einstein-Hilbert\footnote{abbreviated as EH from now on.} Lagrangian) for a classical, stationary black hole. From there we can compute the change in area using the Raychaudhuri equation, and establish the first law. The analysis in this chapter is dynamical.\\
Using the definition of $\bQ$, i.e, equation(\ref{defJ2}), we rewrite equation(\ref{deH1}) as:
\beq
\de H = \int_\Sigma\xi^\al\de\bK_\al + \int_{\dee\Sigma}[\de\bQ - \xi\cd\bTh]
\label{deH2}
\eeq
We assume that $H$ exists for all infinitesimal asymptotic symmetries and that it is independent of the choice of $\xi^\al$. Further when $\xi^\al$ is Killing then it can be shown trivially (equation (23) of \cite{w2}) that $\de H$ is independent of our choice of slice, $\Sigma$.\label{invariant}\\ \\
Now we assume that $\de\phi$ satisfies linearized equations of motion throughout the spacetime. Also we choose our slice (see figure[\ref{surface}]) such that it extends smoothly to the boundary representing infinity. (we are free to choose our slices because of the slice independence of $\de H$) 

Doing so, and remembering our assumption we have $\de \bK_\al \rightarrow 0$, equation(\ref{deH2}) takes the form:
\beq
\de H = \int_{\infty}(\de\bQ[\xi] - \xi\cd\bTh)
\label{deH3}
\eeq
Stokes' theorem is used now, to rewrite the above as: 
\beq
\de H = \int_{\Sigma}(\de d\bQ[\xi] - d(\xi\cd\bTh)) + \int_{\dee\Sigma}(\de\bQ[\xi] - \xi\cd\bTh)
\label{deH4}
\eeq
here, $\dee\Sigma$ is any interior boundary of $\Sigma$. Using equation(\ref{delj}) we eliminate the $d(\xi\cd\bTh)$ term to get:
\beq
\de H = \int_{\Sigma}(\de d\bQ[\xi] - \de\bJ[\xi]) + \int_{\dee\Sigma}(\de\bQ[\xi] - \xi\cd\bTh)
\eeq
Note that we have restricted our attention to $\xi^\al$ Killing, which is why $\Om$ in equation(\ref{delj}) is zero. Using equation(\ref{defJ2}) in the first integral of the above equation, we get:
\beq
\de H = - \int_\Sigma\xi^\al\de\bK_\al + \int_{\dee\Sigma}[\de\bQ - \xi\cd\bTh]
\label{deH5}
\eeq
It is important to note that the above equation does not require $\de\phi$ to satisfy linearized equations of motion throughout 
the spacetime, in that case $\de\bK_\al = 0$ and the integral over $\Sigma$ is zero. Our equation(\ref{deH5}) allows 
\begin{enumerate}
\item The presence of sources for Einstein's equations as well as for other matter fields.
\item $\dee\Sigma$ to be arbitrary.
\end{enumerate}

We look into the EH Lagrangian now, and derive formulae for variation in ADM mass and angular momentum. The EH Lagrangian is:
\beq
\bL = \frac{1}{16\pi}\textbf{\eps}R \text{\quad where, $\textbf\eps$ is the associated volume element}
\label{ehl}
\eeq
The first order variation of $\bL$ gives
\beq
\de\bL = \frac{1}{16\pi}\textbf{\eps}(-R^{\al\be} + \frac{1}{2}g^{\al\be}R)\de g_{\al\be} + d\bTh
\eeq
where, 
\beq
\begin{split}
\bTh_{\al\be\ga}(\phi,\de\phi) = \frac{1}{16\pi}\eps_{\de\al\be\ga}v^\de \\
\text{with,\qquad} v_\de = \del^\be\de g_{\de\be} - g^{\mu\nu}\del_\de \de g_{\mu\nu}
\label{theta}
\end{split}
\eeq
To use equation(\ref{deH5}) to define variations in ADM mass and angular momentum, we need to identify $\bQ$ and $\bK_\al$ which we go on to do next.

%----------identifying Q and K------------
\newpage
\subsection{Identifying $\bQ$ and $\bK_\al$}
We start with equation(\ref{defJ1}) and try to bring it to the form of equation(\ref{defJ2}). \\
We shall use in this subsection the following formulae/definitions : 
\\The Einstein tensor, defined as: \beq
R^{\al\be} - \frac{1}{2}g^{\al\be}R = G^{\al\be}
\label{et}
\eeq
The variation of metric: \beq\de g^{\al\be} = \del^{(\al}\xi^{\be)}
\label{delg}
\eeq
The Ricci identity: \beq\del^{[\al}\del^{\be]}\xi^\ga = R^{\al\be\ga}_{\quad\thickspace\de}\xi^\de
\label{ri}
\eeq

In the case of the EH Lagrangian, with the $\bL$ given by equation(\ref{ehl}) and $\bTh$ identified as equation(\ref{theta}), equation(\ref{defJ1}) can be 
written as:
\beq
\bJ = \frac{1}{16\pi}\eps_{\de\al\be\ga}[(\del_\rho\de g^{\de\rho} - g_{\mu\nu}\del^\de g^{\mu\nu}) - R\xi^\de]
\eeq
Using equation(\ref{delg}) we rewrite above equation as: 
\beq
\bJ = \frac{1}{16\pi}\eps_{\de\al\be\ga}\del_\rho\del^{(\de}\xi^{\rho)} - 
\frac{1}{16\pi}\eps_{\de\al\be\ga}g_{\mu\nu}\del^\de\del^{(\mu}\xi^{\nu)} 
 - \frac{1}{16\pi}\eps_{\de\al\be\ga}R\xi^\de 
\eeq
or, \beq
\bJ = \frac{1}{16\pi}\eps_{\de\al\be\ga}\del_\rho\del^{(\de}\xi^{\rho)} - 
\frac{1}{16\pi}\times 2\times\eps_{\de\al\be\ga}g_{\mu\nu}\del^\de\del^{\mu}\xi^{\nu} 
 - \frac{1}{16\pi}\eps_{\de\al\be\ga}R\xi^\de 
\eeq
or, \beq\begin{split}
\bJ = \frac{1}{16\pi}\eps_{\de\al\be\ga}\del_\rho\del^{(\de}\xi^{\rho)} - 
\frac{1}{16\pi}\eps_{\de\al\be\ga}g_{\mu\nu}[\del^{(\de}\del^{\mu)}\xi^{\nu} \\
+ \del^{[\de}\del^{\mu]}\xi^{\nu}] - \frac{1}{16\pi}\eps_{\de\al\be\ga}R\xi^\de
\end{split}
\eeq
Using equation(\ref{ri}),
\beq
\begin{split}
\bJ = \frac{1}{16\pi}\eps_{\de\al\be\ga}\del_\rho\del^{(\de}\xi^{\rho)} - 
\frac{1}{16\pi}\eps_{\de\al\be\ga}g_{\mu\nu}\del^{(\de}\del^{\mu)}\xi^{\nu} \\
- \frac{1}{16\pi}\eps_{\de\al\be\ga}g_{\mu\nu}R^{\de\mu\nu}_{\quad\thickspace\rho}\xi^\rho - \frac{1}{16\pi}\eps_{\de\al\be\ga}R\xi^\de
\end{split}
\eeq
Now using definition equation(\ref{et}) we have,
\beq
\begin{split}
\bJ = \frac{1}{16\pi}\eps_{\de\al\be\ga}\del_\rho\del^{(\de}\xi^{\rho)} - 
\frac{1}{16\pi}\eps_{\de\al\be\ga}g_{\mu\nu}\del^{(\de}\del^{\mu)}\xi^{\nu} \\
+ \frac{1}{16\pi}\eps_{\de\al\be\ga}(G^\de_\rho + \frac{R}{2}\de^\de_\rho)\xi^\rho - \frac{1}{16\pi}\eps_{\de\al\be\ga}R\xi^\de
\end{split}
\eeq

or, \beq
\begin{split}
\bJ = \frac{1}{16\pi}\eps_{\de\al\be\ga}\del_\rho\del^{(\de}\xi^{\rho)} - 
\frac{1}{16\pi}\eps_{\de\al\be\ga}\del^{(\de}\del^{\mu)}\xi_{\mu} \\
+ \frac{1}{16\pi}\eps_{\de\al\be\ga}G^\de_\rho\xi^\rho - \frac{1}{32\pi}\eps_{\de\al\be\ga}R\xi^\de
\end{split}
\eeq

or, \beq
\begin{split}
\bJ = \frac{1}{16\pi}\eps_{\de\al\be\ga}\del_\rho\del^{(\de}\xi^{\rho)} - 
\frac{1}{16\pi}\eps_{\de\al\be\ga}[2\del^\mu\del^\de-\del^{[\mu}\del^{\de]}]\xi_{\mu} \\
+ \frac{1}{16\pi}\eps_{\de\al\be\ga}G^\de_\rho\xi^\rho - \frac{1}{32\pi}\eps_{\de\al\be\ga}R\xi^\de
\end{split}
\eeq

Using equations(\ref{et}) and (\ref{ri}), \beq
\begin{split}
\bJ = \frac{1}{16\pi}\eps_{\de\al\be\ga}g_{\rho\sigma}\del^\sigma\del^{(\de}\xi^{\rho)} - 
\frac{1}{8\pi}\eps_{\de\al\be\ga}g_{\rho\sigma}\del^\sigma\del^\de\xi^\rho + \frac{1}{16\pi}\eps_{\de\al\be\ga}[G^\de_\rho + \frac{R}{2}\de^\de_\rho]\xi^\rho \\
+ \frac{1}{16\pi}\eps_{\de\al\be\ga}G^\de_\rho\xi^\rho - \frac{1}{32\pi}\eps_{\de\al\be\ga}R\xi^\de
\end{split}
\eeq

or,
\beq
\begin{split}
\bJ = -\frac{1}{16\pi}\eps_{\de\al\be\ga}\del_\rho\del^{[\de}\xi^{\rho]} + \frac{1}{8\pi}\eps_{\de\al\be\ga}G^\de_\rho\xi^\rho	\\
= A_{\al\be\ga} + \frac{1}{8\pi}\eps_{\de\al\be\ga}G^\de_\rho\xi^\rho \label{A}
\end{split}
\eeq
We shall now try to write the first term, $A_{\al\be\ga}$  as $d\bQ$ where, $\bQ$ is a 2-form.\\ Let us call it, $\bQ_{\al\be}$.\\
So,\beq
d\bQ_{\ga\al\be} = \del_{[\ga}\bQ_{\al\be]}
\label{dq}
\eeq
Let us look at the dual of equation(\ref{dq}),
\beq
\begin{split}
\ast(d\bQ)_{\ga\al\be} = \eps^{\mu\ga\al\be}\del_\ga{\bQ}_{\al\be} \\
= \del_\ga\tilde{\bQ}^{\mu\ga}	\label{stardq}\\
\text{where,\qquad} \tilde\bQ^{\mu\ga} = \eps^{\mu\ga\al\be}\bQ_{\al\be}
\end{split}
\eeq
But, from equation(\ref{A}), 

\beq
\begin{split}
\ast A_{\al\be\ga} = \eps^{\mu\al\be\ga}A_{\al\be\ga} \\
= -\frac{1}{16\pi}\eps^{\mu\al\be\ga}\eps_{\de\al\be\ga}\del_\rho\del^{[\de}\xi^{\rho]} \\
= \del_\ga(-\frac{1}{16\pi}\del^{[\mu}\xi^{\ga]}) \\
\label{starA}
\end{split}
\eeq

Comparing equations (\ref{stardq}) and (\ref{starA}) we identify:
\beq
\tilde\bQ^{\mu\ga} = -\frac{1}{16\pi}\del^{[\mu}\xi^{\ga]}
\eeq
So,\beq
\bQ_{\al\be} = -\frac{1}{16\pi}\eps_{\al\be\ga\de}\del^{[\ga}\xi^{\de]}
\label{Q}
\eeq
We have thus cast equation(\ref{defJ1}) into the form of equation(\ref{defJ2}), and identified, 
\beq
\begin{split}
\bK_{\al\be\ga\de} = \frac{1}{8\pi}\eps_{\rho\al\be\ga}G^\rho_\de \\
\bQ_{\al\be} = -\frac{1}{16\pi}\eps_{\al\be\ga\de}\del^{[\ga}\xi^{\de]}
\label{identify}
\end{split}
\eeq
\clearpage

\subsection{First-order variations in ADM mass and angular momentum}
Now, let $g_{\al\be}$ be the solution of the vacuum Einstein equations, and let $\de g_{\al\be}$ be a linearized perturbation which satisfies equations of motion with source $\de T_{\al\be}$. \\
Then from the identification equation(\ref{identify}),\beq
\de\bK_{\al\be\ga\de} = \eps_{\rho\al\be\ga}\de T^\rho_{\quad\de}
\label{delc}
\eeq
Now, we can substitute the above formula into equation(\ref{deH5}),
\beq
\de H = - \int_\Sigma\eps_{\rho\al\be\ga}\xi^\de\de T^\rho_{\quad\de} + \int_{\dee\Sigma}[\de\bQ - \xi\cd\bTh]
\label{deH6}
\eeq
\\ADM mass is defined to be the value of the Hamiltonian for the EH Lagrangian in an asymptotically flat spacetime where the Killing vector corresponds to time translation, $t^\al$. Writing $H = M$ equation(\ref{deH6}) corresponds to:
\beq
\de M = - \int_\Sigma\eps_{\rho\al\be\ga}t^\de\de T^\rho_{\quad\de} + \int_{\dee\Sigma}[\de\bQ[t] - t\cd\bTh]
\label{delM1}
\eeq
Similarly, writing $H = -J$ and choosing $\xi^\al$ as asymptotic rotation, $\varphi^\al$, the variation in angular momentum in an asymptotically flat spacetime is: 
\beq
\de J = \int_\Sigma\eps_{\rho\al\be\ga}\varphi^\de\de T^\rho_{\quad\de} - \int_{\dee\Sigma}[\de\bQ[\varphi] - \varphi\cd\bTh]
\label{delJ1}
\eeq
\clearpage

\subsection{The physical process derivation of the First law}
\begin{figure}[!h]
    \begin{center}
        \scalebox{0.45}{
            \includegraphics[scale=.75]{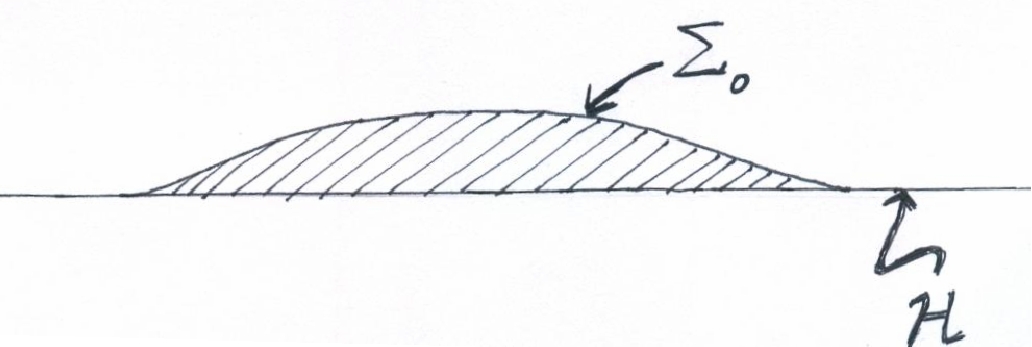}}
    \end{center}
            \caption{\small \textit{Slice} asymptotically terminates on event horizon $\mathcal{H}$, no contribution to $\de H$ comes from the integral over the shaded region}
            \label{approach}
\end{figure}

We consider a classical, stationary black hole solution to the vacuum EH equations. We put in some matter into it, (as a perturbation) and assume that the black hole is not destroyed in the process, but settles down to a stationary final state. Using equations(\ref{delM1}) and (\ref{delJ1}) we can find change in mass and angular momentum. Also Raychaudhuri equation will give us the change in area. We shall see how the change in all these 3 quantities relate to each other. That is precisely the statement of the first law of black hole mechanics.\\
We let $g_{\al\be}$ be the solution to the source free EH equations of motion corresponding to a stationary black hole. Let,
\beq
\xi^\al = t^\al + \Omega_H\varphi^\al
\label{killing}
\eeq
be the Killing field of this black hole. Let $\Sigma _0$ be an asymptotically flat hyperspace which terminates on the event horizon $\mathcal{H}$ of the black hole.(figure[\ref{approach}])

We consider initial data on $\Sigma_0$ for a linearized perturbation with matter source $\de T^\al_{\quad\be}$. As before we require that $\de T^\al_{\quad\be}$ vanish near infinity (so that we are justified to use equation(\ref{deH6}) and hence our definitions of ADM mass and angular momentum). In addition we require that the initial data for $\de g_{\al\be}$ vanish near the neighborhood of the horizon $\mathcal{H}$ on $\Sigma$, so that the 2-form integral arising in equations(\ref{delM1}) and (\ref{delJ1}) vanishes. Combining the definitions of ADM mass and angular momentum, we get: 
\beq
\de M - \Omega_H\de J = - \int_{\Sigma_0}\eps_{\rho\al\be\ga}\xi^\de\de T^\rho_{\quad\de}
\label{fl1}
\eeq
Let $n^\al$ be the unit future-like normal to $\Sigma_0$, then
\beq
- \eps_{\rho\al\be\ga} = n_\rho\tilde{\eps_{\al\be\ga}}
\label{eps}
\eeq
Now, since $\Sigma _0$ terminates on the event horizon $\mathcal{H}$ of the black hole we can replace $n_\rho$ with null tangent $k_\rho$. Making these changes in equation(\ref{fl1}) we have,
\beq
\de M - \Omega_H\de J = \int_\mathcal{H}\Tilde{\eps_{\al\be\ga}}\xi^\de\de T^\rho_{\quad\de}k_\rho = \int_\mathcal{H}\xi^\de\de T^\rho_{\quad\de}k_\rho
\label{fl2}
\eeq
We shall now try to cast the R.H.S of the above equation in terms of change in area. The Raychaudhuri equation for the stationary black hole using Einstein's equations corresponds to:
\beq
\frac{d\theta}{dV} = - 8\pi\de T^\rho_{\quad\de}k_\rho k^\de
\label{rc1}
\eeq
Since we have assumed that the Zeroth Law of black hole mechanics holds, we can substitute for $k^\de$ in the R.H.S of above equation in terms of the constant surface gravity\footnote{Using the relation between affine parameter($V$) and Killing parameter($v$), $ln V = \kappa v$;}
\beq
k^\de = (\frac{\dee}{\dee V})^\de = \frac{1}{\kappa V}\xi^\de \label{affpara}
\eeq
Using equations(\ref{rc1}) and (\ref{affpara}) the integration over the horizon $\mathcal{H}$, [the R.H.S. of equation(\ref{fl2})], gives
\beq
\int_\mathcal{H}\xi^\de\de T^\rho_{\quad\de}k_\rho = - \frac{1}{8\pi}\kappa\int_0^\infty dV\int_{\dee\mathcal{H}} d^2S\;\;V\frac{d\theta}{dV}
\label{fl3}
\eeq
The right hand can be integrated by parts,
\beq
- \frac{1}{8\pi}\kappa\int_{\dee\mathcal{H}} d^2S \int_0^\infty V\frac{d\theta}{dV} dV = - \frac{1}{8\pi}\kappa [ \int_{\dee\mathcal{H}}d^2S\;\;(\theta V) |_0^\infty - \int_{\dee\mathcal{H}}d^2S\;\;\int_0^\infty\theta dV]
\label{areaint}
\eeq
From the interpretation of $\theta$ as $\frac{1}{A}\;\frac{dA}{dV}$, with $A$ as area of the black hole, the second term in equation(\ref{areaint}) is just the change in black hole area. On the other hand, the first term vanishes, since $\theta\rightarrow 0$ as $V\rightarrow 0$ and also as $V\rightarrow\infty$, since the black hole starts from an initial stationary state and settles down to a final stationary state according to our assumption. Thus we obtain,
\beq
- \frac{1}{8\pi}\kappa\int_{\dee\mathcal{H}} d^2S \int_0^\infty V\frac{d\theta}{dV} dV = \frac{1}{8\pi}\kappa\de A
\label{delA}
\eeq
Thus from equations (\ref{fl2}),(\ref{fl3}) and (\ref{delA}) we obtain the First Law of black hole mechanics,
\beq
\de M - \Omega_H\de J = \frac{1}{8\pi}\kappa\de A
\label{fl4}
\eeq
or,
\beq
\de M - \frac{1}{8\pi}\kappa\de A = \Omega_H\de J
\label{fl5}
\eeq
Keeping in mind equation(\ref{fl5}) we note the following analogies,
\begin{description}
\item [1] The first law of thermodynamics, which states $dE - TdS = - \textit{work done}$
\item [2] The result of Hawking radiation, which identifies surface gravity, $\kappa$, with temperature, $T$.
\item [3] The result of second law of black hole \textit{mechanics}, which states that the change in area, $\de A$, is always positive.
\item [4] The second law of thermodynamics, which states that the change in entropy, $\de S$, is always positive.
\end{description}
The above four analogies strongly motivates us to identify area with entropy! Hence we rewrite equation(\ref{fl4}) as:
\beq
\de M - \Omega_H\de J = \frac{\kappa}{2\pi}\de \bS
\label{fl6}
\eeq \\
The above equation is assumed to hold not only for the EH Lagrangian, but for all classical theories of gravity arising from a diffeomorphism invariant Lagrangian. However in other such theories the entropy need not be equal to the area of the black hole. 
\clearpage
\newpage

\section{Identification of Noether charge with entropy}
We shall now show how the formalism developed in the first section of this chapter contains the identification of entropy with the Noether charge (assuming the zeroth and the first law holds). \\

We begin our analysis from equation(\ref{deH3}) which holds for any diffeomorphism invariant theory. In this section we use this formula to define variations in canonical mass and angular momentum. Once again, we call $H = M$ for the choice of $\xi^\al = t^\al$ and $H = - J$ for $\xi^\al = \varphi^\al$. Equation(\ref{deH3}) gives us:
\beq
\de M = \int_{\infty}(\de\bQ[t] - t\cd\bTh)
\label{delM}
\eeq
\beq
\de J = - \int_{\infty}\de\bQ[\varphi]
\label{delJ}
\eeq
In the above equation $\varphi\cd\bTh$ does not appear since $\varphi^\al$ is assumed to be tangent to the ($n$ - 2) dimensional sphere where the integrals are evaluated. Now we specialize to a stationary black hole solution with a bifurcate Killing horizon with a bifurcation surface, $\dee\Sigma$. Once again we choose $\xi^\al$ to be the Killing field which vanishes on $\dee\Sigma$.\footnote{see equation(\ref{killing}).} For the Killing case, the symplectic current $\Om$ vanishes, and equations(\ref{defJ2}) and (\ref{delj}) gives us:
\beq
d(\de\bQ) = d(\xi\cd\bTh)
\eeq
Integrating the above equation over $\Sigma$, taking into account equations(\ref{delM}) and (\ref{delJ}) we obtain:
\beq
\de M - \Om_H\de J = \de\int_{\dee\Sigma}\bQ
\label{wald}
\eeq
However the right hand side of the above equation need to be written as $\kappa$ times something, so that the identification with entropy can be made. The analysis in this section is independent of the Lagrangian, and in this respect it is important that the Zeroth law holds irrespective of the Lagrangian. \\
We follow the algorithm as suggested by Wald\cite{w1}. Define a ($n$ - 2)-form $\tilde\bQ$ on $\dee\Sigma$ by expressing $\bQ$ in terms of $\xi^\al$ and $\del_\al\xi_\be$. Since $\xi^\al$ vanishes on $\dee\Sigma$, and $\del_\al\xi_\be = \kappa\varepsilon_{\al\be}$\footnote{$\varepsilon_{\al\be}$ is the binormal to $\dee\Sigma$}, all references to $\xi^\al$ has been eliminated. Now since $\xi$ and $\kappa$ scales in the same way, if we choose $\tilde\bQ$ to have unit surface gravity, then on $\dee\Sigma$ we have,
\beq
\de\bQ = \kappa\de\tilde\bQ
\eeq
Thus, we can pull out $\kappa$ from equation(\ref{wald})
\beq
\de M - \Om_H\de J = \kappa\;\de\int_{\dee\Sigma}\tilde\bQ
\label{wald1}
\eeq
Comparing equations(\ref{fl6}) and (\ref{wald1}) we see, that black hole entropy, $\bS$ is defined by:
\beq
\bS = 2\pi\int_{\dee\Sigma}\tilde\bQ
\label{wald3}
\eeq
The above formula establishes black hole entropy in terms of Noether charge for a general diffeomorphism invariant classical theory which admits stationary black hole solutions. Note that the identification(\ref{wald3}) is based on the validity of the first law, equation(\ref{fl6}). The first law was derived using the EH Lagrangian. The identification(\ref{wald3}) is hence a definition of entropy which depends on the \textit{dynamics} of the theory!

\chapter{Conclusion}
The main theme of this thesis report has been to arrive at the expression for entropy in the case of  stationary black holes. We started our analysis by deriving the established laws of black hole mechanics. All through our analysis we had been careful to distinguish laws which were consequences of kinematics from the dynamical ones.\\
We saw in chapter 2, that surface gravity characterized equilibrium for stationary black holes. After providing the proof by Bardeen, Carter and Hawking (see section \ref{sec:bchproof}) which makes use of the Einstein's equations (equation(\ref{eq:einstein})), the kinematic proof due to Wald (see section \ref{sec:zerothwald}) has been discussed. However this proof relies on the existence of a bifurcation surface. In chapter 3, we looked at the modes of a scalar field in the presence of a spherically symmetric background metric. A straightforward calculation led us to identify surface gravity as Hawking temperature (equation(\ref{kappaisT})). Nowhere in the calculation were the equations of motion used. Thus once again kinematics showed that temperature and surface gravity are the same quantity for stationary black holes (in equilibrium). This indicates (in analogy to the Zeroth law of thermodynamics) why we should expect the uniformity of surface gravity over the event horizon to be valid kinematically.\\
Next we dealt with the second law of black hole mechanics in chapter 4. Once we used the equations of motion and posed certain energy conditions, the Raychaudhuri equation showed that the area of the event horizon of a stationary black holes can never decrease (equation(\ref{areath})). It is here that the dynamics starts to play a role.\\
In chapter 5 we developed Wald's formulation and derived the `physical' process version of the first law of black hole mechanics for the Einstein-Hilbert Lagrangian. The first law gave us a relation between \textit{work done}, \textit{mass} and the \textit{area} of the event horizon for a stationary black hole (equation(\ref{fl5})). Then we used all the laws of black hole mechanics to motivate us to identify entropy with area for the Einstein-Hilbert Lagrangian (equation(\ref{fl6})). Wald's formulation was generalized to arbitrary diffeomorphism invariant Lagrangians and a equation equivalent to the first law was obtained (equation(\ref{wald1})). This allowed us to define entropy for any theory of gravity. The entropy so defined turned out to be a purely \textit{geometric} quantity (equation(\ref{wald3})). It is the surface integral of the \textit{Noether charge} of the diffeomorphism invariance current which depends only on the dynamical fields appearing in the Lagrangian. However we must note that we needed to assume that the Killing horizon is bifurcate.\footnote{In our derivation of first law for Einstein-Hilbert Lagrangian, there were no such assumptions.} This analysis fails for extremal black holes where the horizon is not a bifurcate. \\ \\ In the following table we summarize the analogy between Einstein-Hilbert black holes' mechanics and laws of thermodynamics: \\

\begin{table}[h!]
\begin{center}
\begin{tabular}{ | l | p{5cm} | p{5cm}| }
  \hline \hline
\textbf{Law} & \textbf{Thermodynamic system} & \textbf{Black hole} \\
\hline
Zeroth law & $T$ constant on a body in thermal equilibrium & $\kappa$ constant over a stationary black hole's event horizon \\
First law & $dE = TdS - PdV$ & $\de M = \frac{\kappa}{8\pi} \de A + \Omega \de J$\\
Second law & $\de S \geq 0$ & $\de A \geq 0$ \\
\hline
\end{tabular}
\caption{Analogy between laws of thermodynamics and laws of black hole mechanics}
\end{center}
\end{table}
\newpage It is interesting to note that the dependence of entropy on the dynamics puts strong constraints in the Lagrangian when one demands the Second law to be respected. For instance if one takes, a higher dimensional theory of gravity of the form:
\beq
I_0 = \int d^Dx \frac{1}{16\pi G}\sqrt{-g}(R + P(R))
\label{hdgravity}
\eeq
In the above expression for action, $P(R)$ is a polynomial in the Ricci scalar $R$. Now using the Wald formulation the entropy can be computed$^{\cite{ted}}$. It is given by, 
\beq
S(g) = \frac{1}{4G}\int_\mathcal{H}d^{D-2}x\sqrt{h}(1+P'(R))
\eeq
Second law now restricts the coefficients in $P(R)$ by ensuring that $1 + P'(R)$ remains positive everywhere outside and on the event horizon of the black hole spacetime. This is a very good example which shows the Second law puts constraints on the theory.\\ 
In this thesis we have posed and answered some questions related to black hole thermodynamics. \textit{What are the precise inputs that go into each of the laws of black hole mechanics?} \textit{Which of these are kinematical and which are dynamical?} To answer these questions we have collected and summerized results which are scattered in the literature. It is hoped that others seeking an introduction to this fascinating subject will find this thesis helpful. 
In the process of answering these questions, we have understood the assumptions which are really necessary. This inevitably leads to further questions. Wald's formulation of geometric entropy gives a general perspective on black hole entropy for any dimension and any Lagrangian. Any future quantum theory of gravity will be tested on its ability to reproduce Wald's formula in the appropriate classical limit.

\appendix
%\addcontentsline{toc}{chapter}{\APPENDICES}
\chapter{The Raychaudhuri equation}

Let $\mo$ be an open region in spacetime. A $congruence$ in $\mo$ is a family of curves such that through each point in $\mo$ there passes one and only one curve from this family. (picture this as a bundle of non-intersecting wires.) In this appendix we will be interested in the evolution of such a congruence. That is precisely what Raychaudhuri equation tells us. 

\section{Newtonian derivation of Raychaudhuri equation}
We present first a Newtonian derivation of the Raychaudhuri equation which captures the essence in the timelike case. \\ \\
We consider a pressureless fluid with velocity given by $\vec{v}(x,t)$. Now if we have any scalar function $f$ then we define the convective derivative as,
\beq
\frac{df}{dt} = \frac{\dee f}{\dee t} + \vec{v}\cd\vec{\del}f
\label{diff1}
\eeq
Also, we define $expansion$ as the divergence of $\vec{v}$ and call it $\theta = \vec{\del}\cd\vec{v}$. Our programme is to find the time-evolution of $\theta$. On differentiating $\theta$ with $t$,
\begin{eqnarray}
\frac{d\theta}{dt} = \frac{\dee \theta}{\dee t} + \vec{v}\cd\vec{\del}\theta \\
or, \frac{d\theta}{dt} = \dee_i\dee_t v^i + v^j\dee_i\dee_j v^i \\
= \dee_i\dee_t v^i + \dee_i{v^j\dee_j v^i} - (\dee_iv^j)(\dee_j v^i) \\
= \dee_i (\dee_t v^i + v^j\dee_j v^i) - (\dee_iv^j)(\dee_j v^i) \label{diff2}
\end{eqnarray}
Using equation(\ref{diff1}) we identify the term within parentheses in equation(\ref{diff2}) as \beq \dee_t v^i + v^j\dee_j v^i = \frac{dv}{dt}\eeq The second term in equation(\ref{diff2}) can be identified with a second rank tensor. Any tensor can be decomposed into its symmetric and antisymmetric (which we call $\omega_{ij}$) parts. The symmetric part can be further decomposed into the trace and the traceless (we call it $\sigma_{ij}$) parts. Since the trace $\dee_i v_i$ is just $\theta$ by definition, we have: 
\beq
(\dee_iv^j)(\dee_j v^i) = \frac{\theta^2}{3} + \sigma^2 - \omega^2
\label{decom}
\eeq
In our case $\omega_{ij} = \dee_{[i}v_{j]}$ corresponds to rotation and $\sigma_{ij} = \dee_{(i}v_{j)} - \dee_{i}v_{i}$ corresponds to the shear. The factor of 3 in the first term of equation(\ref{decom}) comes since our matrices are all of $3\times 3$ dimensions.
Now we re-write equation(\ref{diff2}) as:
\beq
\frac{d\theta}{dt} =  \vec\del\cd\frac{d\vec v}{dt} - \frac{\theta^2}{3} - \sigma^2 + \omega^2 \label{diff3}
\eeq
The first term on the R.H.S. of equation(\ref{diff3}) is just the divergence of acceleration. Since we are in a theory of gravity (Newtonian) where force is conserved, we may express acceleration in terms of the Newtonian potential:
\beq
\frac{d\vec v}{dt} = -\vec\del\Phi_{gr}
\label{acc}
\eeq
If we have a mass density($\rho$) as source, then $\Phi_{gr}$ satisfies Poisson's equations:
\beq
\del^2\Phi_{gr} = 4\pi G\rho
\label{pe}
\eeq
Substituting for the first term in equation(\ref{diff3}) using equations(\ref{acc}) and (\ref{pe}) we get the Raychaudhuri equation,
\beq
\frac{d\theta}{dt} = -4\pi G\rho - \frac{\theta^2}{3} - \sigma^2 + \omega^2 \label{rc1}
\eeq

\newpage
\section{General derivation of the Raychaudhuri equation}
In analogy to the Newtonian derivation, we find the evolution of the $expansion$ $scalar$, $\theta$ obtained by taking the trace of tensor field \beq B_{\al\be} = u_{\al ;\be}\label{B}\eeq where \beq u^\al = \frac{dx^\al}{d\tau}\eeq is the tangent to the geodesic of the congruence. We develop the equation for $\theta$ by finding an equation for $\be_{\al\be}$.
\bea
B_{\al\be ;\mu}u^\mu = u_{\al ;\be\mu} u^\mu \\
= (u_{\al ;\mu\be} - R_{\al\nu\be\mu}u^\nu) u^\mu \\
= -B_{\al\mu}B^\mu_\be - R_{\al\mu\be\nu}u^\mu u^\nu
\eea
On taking the trace of the above equation,
\beq
\frac{d\theta}{d\tau} = -B^{\al\be}B_{\be\al} - R_{\al\be}u^\al u^\be
\label{rc2}
\eeq
It is easy to see that for $u^\al$ = timelike,
\beq
\frac{d\theta}{d\tau} = - \frac{\theta^2}{3} - \sigma^2 + \omega^2 - R_{\al\be}u^\al u^\be
\label{rctime}
\eeq
where, $\sigma$ and $\omega$ have the same interpretation as in equation(\ref{decom})\footnote{Note that $B_{\al\be}$ is purely transverse, thus the interpretation of $\sigma$ and $\omega$ goes through.}. Also in the Newtonian limit equation(\ref{rctime}) goes over to equation(\ref{rc1}).\\
There is some subtlety however for $u^\al$ = null. Since $u^\al u_\al = 0$ does not imply that vector $B^{\al\be}$ is tranverse to $u^\al$.\footnote{The transverse space is two dimensional. So we make replacements : $\al\rightarrow A$ and $\be\rightarrow B$} Since the tensor $B_{AB}$ is not purely transverse, we construct the purely transverse tensor $\tilde{B}_{AB}$ by taking projections with the purely transverse induced metric $\ga_{AB}$, which in this case is 2-dimensional. Then we can decompose,
\beq
\tilde{B}_{\al\be} = \frac{\theta}{2}\ga_{AB} + \sigma_{AB} + \omega_{AB}
\label{decom2}
\eeq
It can be checked that \beq \tilde{B}^{BA}\tilde{B}_{AB} = B^{BA}B_{AB} \eeq which means we can rewrite equation(\ref{rc2}) for the null case as: \beq 
\frac{d\theta}{d\tau} = -\tilde{B}^{AB}\tilde{B}_{BA} - R_{\al\be}u^\al u^\be
\label{rc3}
\eeq
And using equation(\ref{decom2}), the final form of Raychaudhuri equation for congruence of null geodesics reads
\beq
\frac{d\theta}{d\tau} = - \frac{\theta^2}{2} - \sigma^2 + \omega^2 - R_{\al\be}u^\al u^\be
\label{rcnull}
\eeq
\newpage
\chapter{Frobenius' Theorem}Congruences which are \textit{hypersurface orthogonal}\footnote{meaning that the congruences are everywhere orthogonal to a family of hypersurface foliating $\mo$, an open region in spacetime} have vanishing rotation tensor. This is the statement of Frobenius' theorem, which we now prove.\\ \\ Suppose that the hypersurfaces are described by equations of the form, $\Phi(x^\al) = c$, where $c$ is a constant specific to each hypersurface. Then the normal to the hypersurface, $n_\al = \Phi_{,\al}$ and since congruences $u^\al$ are orthogonal to them,\beq u_\al = \mu\Phi_{,\al}\label{frob1}\eeq for some proportionality factor $\mu$.\\ Differentiating equation(\ref{frob1}) gives, \beq u_{\al;\be} = \mu_{,\be}\Phi_{,\al} + \mu\Phi_{;\al\be} \label{frob2}\eeq Consider now the antisymmetric tensor, $u_{[\al;\be}u_{\ga]}$. Computation of it using equation(\ref{frob2}) and $\Phi_{;\al\be} = \Phi_{;\be\al}$ gives us zero. Therefore we have the result:\beq\text{hypersurface orthogonal} \implies u_{[\al;\be}u_{\ga]} = 0 \label{frob3}\eeq The converse of the above statement can also be proved to be true. So a congruence of curves (timelike, spacelike, or null) is hypersurface orthogonal if and only if $u_{[\al;\be}u_{\ga]} = 0$, where $u^\al$ is tangent to the curves. \\
Now we focus on 
\bea 
3! u_{[\al;\be}u_{\ga]} = 2(u_{[\al;\be]}u_{\ga} + u_{[\ga;\al]}u_{\be} + u_{[\be;\ga]}u_{\al}) \\ 
= 2(B_{[\al \be]}u_{\ga} + B_{[\ga \al]}u_{\be} + B_{[\be \ga]}u_{\al})\label{frob4} \eea 
By statement(\ref{frob3}) the L.H.S of the above equation gives us zero if geodesics are hypersurface orthogonal.\\ \\For timelike case transvecting the R.H.S of equation(\ref{frob4}) by $u^\ga$ gives $B_{[\al \be]} = 0$, which means by our definition that the rotation tensor vanishes. \\For the null case, transvection with $N^\ga$ gives the desired result.

\newpage
\pagecolor{black}
\includegraphics[width=160mm]{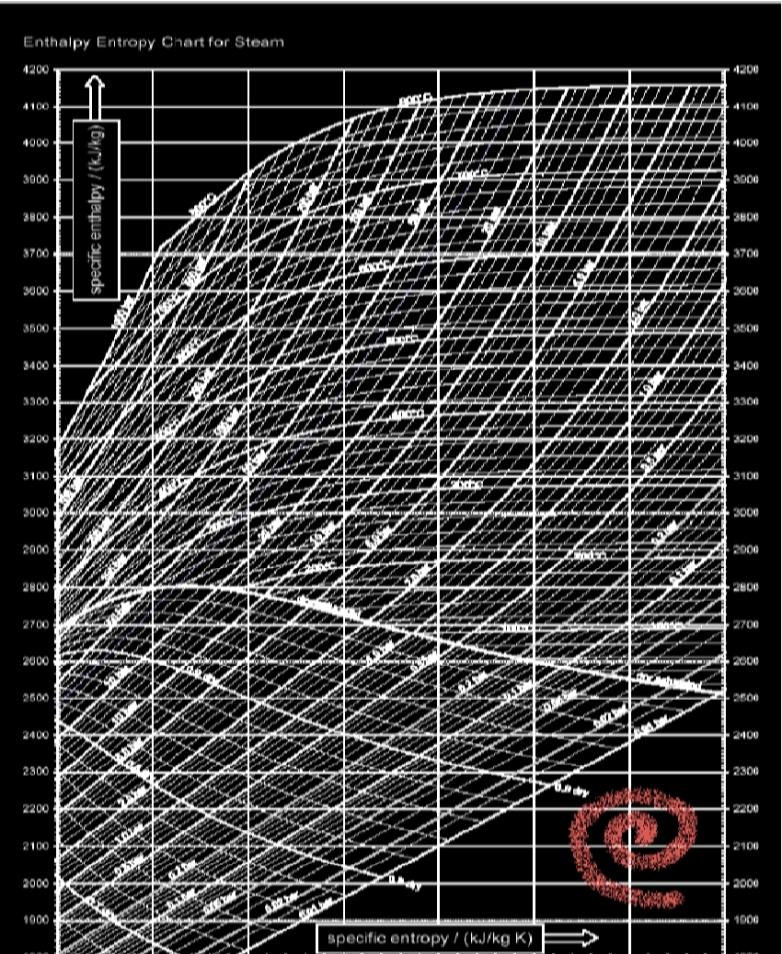}

\end{document}